\documentclass[sigconf]{acmart}
\renewcommand\footnotetextcopyrightpermission[1]{} % removes footnote with conference information in first column
\usepackage{graphicx}
\graphicspath{{figures/}}
\DeclareGraphicsExtensions{.pdf,.jpeg,.jpg,.png}
  
\usepackage{amsmath}
\usepackage{amssymb}
\usepackage{algorithm}
\usepackage{algorithmic}
\usepackage[caption=false,font=footnotesize]{subfig}
\usepackage{url}

\newcommand{\vertexlabel}{\mathit{l}}

\newcommand{\INDSTATE}[1][1]{\STATE\hspace{#1\algorithmicindent}}

\newcommand{\Q}{$\mathcal{Q}$}

\newcounter{saveenumerate}
\makeatletter
\newcommand{\enumeratext}[1]{%
\setcounter{saveenumerate}{\value{enum\romannumeral\the\@enumdepth}}
\end{enumerate}
#1
\begin{enumerate}
\setcounter{enum\romannumeral\the\@enumdepth}{\value{saveenumerate}}%
}
\makeatother

\makeatletter
\newcommand\footnoteref[1]{\protected@xdef\@thefnmark{\ref{#1}}\@footnotemark}
\makeatother

\usepackage{booktabs} % For formal tables

\begin{document}

\title{Loom: Query-aware Partitioning of Online Graphs}

% 1st. author
\author{Hugo Firth}
\affiliation{%
  \institution{School of Computing Science}
  \streetaddress{Newcastle University}
}
\email{h.firth@ncl.ac.uk}
% 2nd. author
\author{Paolo Missier}
\affiliation{%
  \institution{School of Computing Science}
  \streetaddress{Newcastle University}
}
\email{paolo.missier@ncl.ac.uk}
% 3rd. author
\author{Jack Aiston}
\affiliation{%
  \institution{School of Computing Science}
  \streetaddress{Newcastle University}
}
\email{j.aiston@ncl.ac.uk}

\begin{abstract}
%Partitioning large graphs over a cluster of machines improves the scalability of
%a graph processing system, but incurs additional networking cost during
%query execution, due to inter-partition traversals. Query-agnostic
%partitioning algorithms typically minimise the likelihood of any edge
%crossing partition boundaries. These partitionings, however, will be
%sub-optimal with respect to some query workloads, where the queries
%require more frequent traversal of specific subsets of inter-partition
%edges. Furthermore, they are not well suited to operate incrementally on
%dynamic, growing graphs.

As with general graph processing systems, partitioning data over a cluster of machines improves the scalability of graph
database management systems. However, these systems will incur additional network cost during the execution of a query 
workload, due to inter-partition traversals. Workload-agnostic partitioning algorithms typically minimise the likelihood
of any edge crossing partition boundaries. However, these partitioners are sub-optimal with respect to many workloads, 
especially queries, which may require more frequent traversal of specific subsets of inter-partition edges. Furthermore,
they largely unsuited to operating incrementally on dynamic, growing graphs.

We present a new graph partitioning algorithm, Loom, that operates on a stream of graph updates and continuously 
allocates the new vertices and edges to partitions, taking into account a query workload of graph pattern expressions
along with their relative frequencies.
%Our approach starts with the efficient indexing of query patterns into a
	%generalised trie data structure, capturing common patterns of edge
	%traversals when executing queries, which we call motifs. We then compare
	%sub-graphs, which present themselves incrementally in the graph update
	%stream, against these motifs, then attempt to allocate each match to a
	%single partition. This approach reduces the number of inter-partition
	%edges within frequently traversed sub-graphs, improving average query
	%performance. 
%We present an extensive evaluation of the approach over several large test
	%graphs with realistic query workloads and various orderings of the graph
	%updates. We demonstrate that, given a workload, our Loom prototype
	%produces partitionings of significantly better quality than existing
	%streaming graph partitioning algorithms Fennel \& LDG
First we capture the most common patterns of edge traversals which occur when executing queries. We then compare 
sub-graphs, which present themselves incrementally in the graph update stream, against these common patterns. Finally we
attempt to allocate each match to single partitions, reducing the number of inter-partition edges within
frequently traversed sub-graphs and improving average query performance.

Loom is extensively evaluated over several large test graphs with realistic query workloads and 
various orderings of the graph updates. We demonstrate that, given a workload, our prototype produces 
partitionings of significantly better quality than existing streaming graph partitioning algorithms Fennel \& LDG.
\end{abstract}

\maketitle

\section{Introduction} \label{section:introduction}

Subgraph pattern matching is a class of operation fundamental to many
``real-time'' applications of graph data. For example, in social networks
\cite{Gupta2014}, and network security \cite{Choudhury2015a}. Answering a
subgraph pattern matching query usually involves exploring the subgraphs of a
large, labelled graph $G$ then finding those which match a small labelled graph
$q$. Fig.1 shows an example graph $G$ and a set of query graphs $Q$ which we
will refer to throughout.
%Efficiently \textbf{improving the performance of such query graphs} is the primary outcome of this work.
\textbf{Efficiently partitioning large, growing graphs to optimise for given workloads  of such queries} is the primary contribution of this work.
%As an iterative, streaming partitioner, Loom is able to continually
%partition a graph which is growing with time $G\rightarrow G'$. As with

\begin{figure}
	\centering
	\includegraphics[width=\columnwidth]{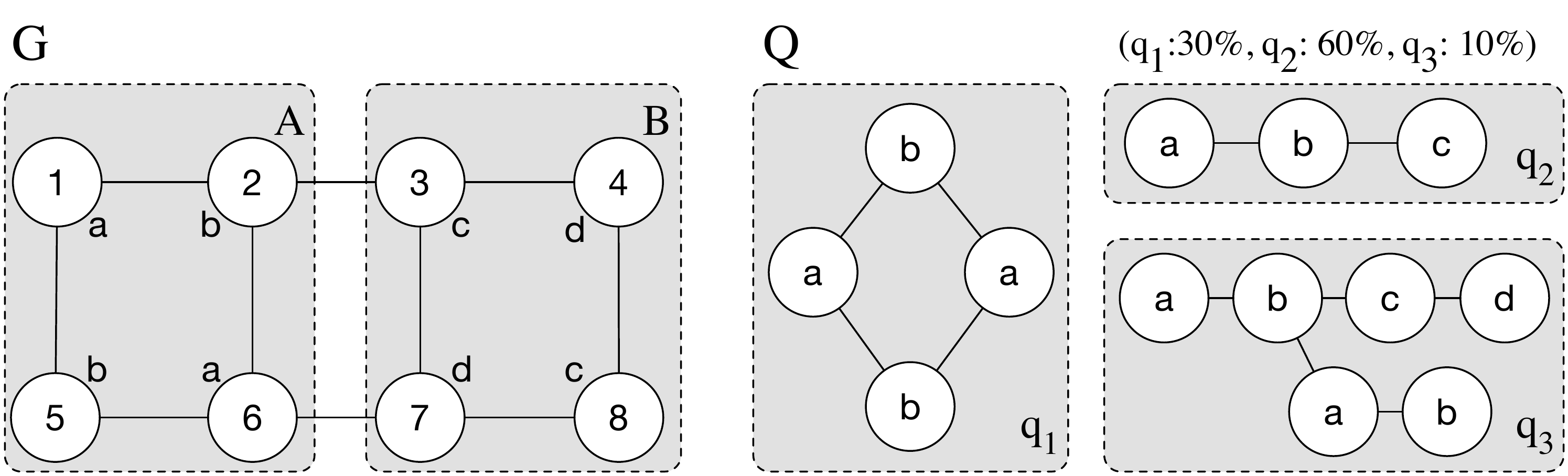}
	\caption{Example graph $G$ with query workload \Q{}}
	\label{fig:graph_plus_workload}	
\end{figure} 

In specialised graph database management systems (GDBMS), pattern
matching queries are highly efficient. They usually correspond to some
index lookup and subsequent traversal of a small number of graph edges,
where edge traversal is analogous to pointer dereferencing. However, as
graphs like social networks may be both large and continually growing, 
eventually they saturate the memory of a single
commodity machine and must be partitioned and distributed. In such a
distributed setting, queries which require inter-partition traversals,
such as $q_2$~in Fig.~\ref{fig:graph_plus_workload}, incur network communication costs and will
perform poorly. A widely recognised approach to addressing these
scalability issues in graph data management is to use one of several
\emph{k-way balanced graph partitioners} \cite{Huang2016, Karypis1998a, Tsourakakis2012, Stanton2012, Xu2014, Margo15, Chevalier2008a}. These systems
distribute vertices, edges and queries evenly across several
machines, seeking to optimise some global goal, e.g.~minimising the
number of edges which connect vertices in different partitions (a.k.a
\emph{min. edge-cut}). In so doing, they improve the performance of a
broad range of possible analyses.

Whilst graphs partitioned for such global measures mostly work well for
global forms of graph analysis (e.g.~Pagerank), no one
measure is optimal for all types of operation \cite{Shang2013}. In
particular, the workloads of pattern matching query workloads, common to GDBMS,
are a poor match for these kinds of partitioned graphs, which we call
\emph{workload agnostic}. This is because, intuitively, a min. edge-cut
partitioning is equivalent to assuming uniform, or at least constant,
likelihood of traversal for each edge throughout query processing. This
assumption is unrealistic as a query workload may traverse a limited subset of
edges and edge types, which is specific to its graph patterns and subject to change.

To appreciate the importance of a workload-sensitive partitioning, consider the graph of Fig.1. 
The partitioning $\{ A, B\} $ is optimal
for the balanced min. edge-cut goal, but may not be optimal for
the query graphs in $Q$. For example, the query graph
$q_2$ matches the subgraphs $\{ (1,2), (2,3)\}$ and $\{ (6,2), (2,3) \}$
in $G$. Given a workload which consisted entirely of $q_2$ queries,
every one would require a potentially expensive inter-partition
traversal (\emph{ipt}). It is easy to see that the alternative
partitioning $A'= \{1,2,3,6\}$, $B' = \{4,5,7,8\}$ offers an
improvement (0 \emph{ipt}) given such a workload, whilst being strictly
worse \emph{w.r.t} min. edge-cut.

Mature research of workload-sensitive online database partitioning is
largely confined to relational DBMS \cite{Quamar2013, Curino2010, Pavlo2012}

\subsection{Contributions}

Given the above motivation, we present Loom: 
%a streaming graph partitioner 
a partitioner for online, dynamic graphs 
which optimises vertex placement to improve the performance
of a given stream of sub-graph pattern matching queries.

The simple goals of Loom are threefold: a) to discover patterns of edge
traversals which are common when answering queries from our given
workload $Q$; b) to efficiently detect instances of these patterns in
the ongoing stream of graph updates which constitutes an online graph; and 
c) to assign these pattern matches wholly within an individual
partition or across as few partitions as possible, thereby reducing the
number of \emph{ipt} and increasing the average performance of any
$q\in Q$.

This work extends an earlier ``vision'' work~\cite{Firth2016c} by the authors,
providing the following additional contributions:

\begin{itemize}
\itemsep1pt\parskip0pt\parsep0pt
\item
	A compact~\footnote{Grows with the size of query graph patterns, which
		are typically small} trie based encoding of the most frequent
		traversal patterns over edges in $G$. We show how it may be
		constructed and updated given an evolving workload $Q$.
\item
  A method of sub-graph isomorphism checking, extending a recent
		probabilistic technique\cite{Song2014}. We show how this measure may be efficiently
  computed and demonstrate both the low probability of false positives
  and the impossibility of false negatives.
\item
  A method for efficiently computing matches for our frequent traversal
  patterns in a graph stream, using our trie encoding and isomorphism
  method, and then assign these matching sub-graphs to graph partitions,
  using heuristics to preserve balance. Resulting partitions do not rely upon
		replication and are therefore agnostic to the complex
		replication schemes often used in
		production systems.
	
\end{itemize}

As online graphs are equivalent to graph streams, we present an extensive
evaluation comparing Loom to popular streaming graph partitioners Fennel
\cite{Tsourakakis2012} and LDG\cite{Stanton2012}. We partition real and
synthetic graph streams of various sizes and with three distinct stream
orderings: breadth-first, depth-first and random order. Subsequently, we execute
query workloads over each graph, counting the number of expensive $ipt$ which
occur. Our results indicate that Loom achieves a significant improvement over
both systems, with between 15 and $40\%$ fewer $ipt$ when executing a given workload.

%
%We present an extensive evaluation of the Loom partitioner, using real and
%synthetic graph streams of various sizes and with three distinct stream
%orderings: breadth-first, depth-first and random order. We subsequently execute
%query workloads over each graph, comparing Loom's performance improvement
%against that of popular streaming graph partitioners Fennel
%\cite{Tsourakakis2012} and LDG\cite{Stanton2012}. Our results indicate that Loom
%achieves a significant improvement over both systems, with between 15 and $40\%$
%fewer inter-partition traversals when executing a given workload.

\subsection{Related work}~\label{subsection:related_work}

Partitioning graphs into k~balanced subgraphs is clearly of practical
importance to any application with large amounts of graph structured
data. As a result, despite the fact that the problem is known to be
NP-Hard \cite{Andreev2006}, many different solutions have been proposed~\cite{Huang2016, Karypis1998a, Tsourakakis2012, Stanton2012, Xu2014, Margo15, Chevalier2008a}. 
We classify these partitioning approaches into one of three potentially overlapping categories: streaming, non-streaming and workload sensitive.
Loom is both a streaming and workload-sensitive partitioner.

\textbf{Non-streaming} graph partitioners \cite{Karypis1998a, Chevalier2008a, Margo15} typically seek
to optimise an objective function global to the graph, e.g.~minimising
the number of edges which connect vertices in different partitions
(\emph{min. edge-cut}).~

A common class of these techniques is known as multi-level partitioners
\cite{Karypis1998a, Chevalier2008a}. These partitioners work by computing a succession of
recursively compressed graphs, tracking exactly how the graph was
compressed at each step, then trivially partitioning the smallest graph
with some existing technique. Using the knowledge of how each compressed
graph was produced from the previous one, this initial partitioning is then
``projected'' back onto the original graph, using a local refinement
technique (such as Kernighan-Lin \cite{Kernighan1970}) to improve the
partitioning after each step. A well known example of a multilevel
partitioner is METIS \cite{Karypis1998a}, which is able to produce high
quality partitionings for small and medium graphs, but performance
suffers significantly in the presence of large graphs \cite{Tsourakakis2012}.
Other mutlilevel techniques \cite{Chevalier2008a} share broadly similar
properties and performance, though they differ in the method used to
compress the graph being partitioned.

Other types of non-streaming partitioner include Sheep \cite{Margo15}: a
graph partitioner which creates an elimination tree from a distributed
graph using a map-reduce procedure, then partitions the tree and
subsequently translates it into a partitioning of the original graph.
Sheep optimises for another global objective function: minimising the
number of different partitions in which a given vertex $v$ has
neighbours (\emph{min. communication volume}).

These non-streaming graph partitioners suffer from two main drawbacks.
Firstly, due to their computational complexity and high memory
usage\cite{Stanton2012}, they are only suitable as offline operations,
typically performed ahead of analytical workloads. Even those
partitioners which are distributed to improve scalability, such as Sheep
or the parallel implementation of Metis (ParMetis) \cite{Karypis1998a}, make
strong assumptions about the availability of global graph information.
As a result they may require periodic re-execution, i.e.~given a dynamic
graph following a series of graph updates, which is impractical online
\cite{Jindal2012}. Secondly, as mentioned, partitioners which optimise for such global measures assume uniform and constant usage of a graph, causing them to ``leave performance on the table'' for many workloads.

\textbf{Streaming} graph partitioners \cite{Tsourakakis2012, Stanton2012, Huang2016} have been proposed to address some of the problems with partitioners outlined above.
Firstly, the strict streaming model considers each element of a graph
stream as it arrives, efficiently assigning it to a partition.
Additionally, streaming partitioners do not perform any refinement,
i.e.~later reassigning graph elements to other partitions, nor do they
perform any sort of global introspection, such as spectral analysis. 
As a result, the memory usage of streaming partitioners
is both low and independent of the size of the graph being partitioned,
allowing streaming partitioners to scale to to very large graphs
(e.g.~billions of edges). Secondly, streaming partitioners may trivially
be applied to continuously growing graphs, where each new edge or update
is an element in the stream.

Streaming partitioners, such as Fennel \cite{Tsourakakis2012} and LDG
\cite{Stanton2012}, make partition assignment decisions on the basis of
inexpensive heuristics which consider the local neighbourhood of each
new element at the time it arrives. For instance, LDG assigns vertices
to the partitions where they have the most neighbours, but penalises
that number of neighbours for each partition by how full it is,
maintaining balance. By using the local neighbourhood of a graph element
$e$ \textbf{at the time} $\mathbf{e}$ \textbf{is added}, such heuristics
render themselves sensitive to the ordering of a graph stream. For
example, a graph which is streamed in the order of a \emph{breadth-first
traversal} of its edges will produce a better quality partitioning than a
graph which is streamed in random order, which has been shown to be
pseudo adversarial\cite{Tsourakakis2012}.

In general, streaming algorithms produce partitionings of lower quality
than their non-streaming counterparts but with much improved
performance. However, some systems, such as the graph partitioner
Leopard \cite{Huang2016}, attempt to strike a balance between the two.
Leopard relies upon a streaming algorithm (Fennel) for the initial
placement of vertices but drops the ``one-pass'' requirement and
repeatedly considers vertices for reassignment; improving quality over
time for dynamic graphs, but at the cost of some scalability.
Note that these Streaming partitioners, like their non-streaming counterparts, are workload agnostic and so share those disadvantages.

\textbf{Workload sensitive} partitioners \cite{Peng2016, Xu2014, Shang2013, Curino2010, Quamar2013, Pavlo2012} attempt to optimise the placement of data to suit a particular workload. 
Such systems may be streaming or non-streaming, but are discussed separately
here because they pertain most closely to the work we do with Loom.

Some partitioners, such as LogGP \cite{Xu2014} and CatchW
\cite{Shang2013}, are focused on improving graph analytical workloads
designed for the \emph{bulk synchronous parallel} (BSP) model of
computation\footnote{e.g.~Pagerank executed using the Apache Giraph
  framework: \url{http://bit.ly/2eNVCnv}.}. In the BSP model a graph processing job is
performed in a number of supersteps, synchronised between partitions.
CatchW examines several common categories of graph analytical workload
and proposes techniques for predicting the set of edges likely to be
traversed in the next superstep, given the category of workload and
edges traversed in the previous one. CatchW then moves a small number of
these predicted edges between supersteps, minimising inter-partition
communication. LogGP uses a similar log of activity from previous
supersteps to construct a hypergraph where vertices which are frequently
accessed together are connected. LogGP then partitions this hypergraph
to suggest placement of vertices, reducing the analytical job's
execution time in future.

%Talk about Peng at als reliance upon replication
In the domain of \textbf{RDF stores}, Peng et al. \cite{Peng2016} use frequent
sub-graph mining ahead of time to select a set of patterns common to a
provided SPARQL query workload. They then propose partitioning
strategies which ensure that any data matching one of these frequent patterns
is allocated wholly within a single partition, thus reducing average
query response time at the cost of having to replicate (potentially
many) sub-graphs which form part of multiple frequent patterns.

For \textbf{RDBMS}, systems such as Schism \cite{Curino2010} and SWORD
\cite{Quamar2013} capture query workload samples ahead of time, modelling
them as hypergraphs where edges correspond to sets of records which are
involved in the same transaction. These graphs are then partitioned
using existing non-streaming techniques (METIS) to achieve a min.
edge-cut. When mapped back to the original database, this partitioning
represents an arrangement of records which causes a minimal number of
transactions in the captured workload to be distributed. Other systems,
such as Horticulture \cite{Pavlo2012}, rely upon a function to estimate
the cost of executing a sample workload over a database and subsequently
explore a large space of possible candidate partitionings.
In addition to a high upfront cost~\cite{Curino2010, Pavlo2012}, these
techniques focus on the relational data model, and so make simplifying
assupmtions, such as ignoring queries which traverse $> 1$-2
edges~\cite{Quamar2013} (i.e. which perform nested joins). Larger traversals are
common to sub-graph pattern matching queries, therefore its unclear how these
techniques would perform given such a workload.

Overall, the works reviewed above either focus on different types of workload
than we do with Loom (namely offline analytical or relational queries), or they
make extensive use of replication. Loom does not use any form of replication,
both to avoid potentially significant storage overheads~\cite{Pujol2010a} and to
remain interoperable with the sophisticated replication schemes used in
production systems.

\subsection{Definitions}\label{section:preliminaries}
Here we review and define important concepts used throughout the rest of the paper.

A \textbf{labelled graph} $G = (V, E, L_V, f_l)$ is of the form: a set
of vertices $V=\{v_1,v_2,...,v_n\}$, a set of pairwise relationships
called edges $e=(v_i,v_j)\in E$ and a set of vertex labels $L_V$. The
function $f_\vertexlabel: V \rightarrow L_V$ is a surjective mapping of
vertices to labels. 
%
%We view a \textbf{graph stream} simply as a
%(possibly infinite) sequence of edges which are being accumulated to a
%graph $G$, over time. We consider \emph{fixed width} sliding windows over such a
%graph stream, i.e.~a sliding window of time $t$
%is equivalent to the $t$ most recent edges in a stream.
%
We view an \textbf{online graph} simply as a (possibly infinite) sequence of
edges which are being added to a graph $G$, over time. We consider \emph{fixed
width} sliding windows over such a graph, i.e.~a sliding window of time $t$ is
equivalent to the $t$ most recently added edges. Note that an online may be
viewed as a \textbf{graph stream} and we use the two terms interchangeably. 

A \textbf{pattern matching query} is defined in terms of sub-graph
isomorphism. Given a pattern graph $q = (V_q, E_q)$, a query should
return $R$: a set of sub-graphs of $G$. For each
returned sub-graph $R_i = (V_{R_i}, E_{R_i})$ there should exist a
bijective function $f$ such that: \emph{a)} for every vertex
$v \in V_{R_i}$, there exists a corresponding vertex $f(v)\in V_q$; \emph{b)}
for every edge $(v_1,v_2)\in E_{R_i}$, there exists a corresponding edge
$(f(v_1),f(v_2))\in E_q$; and \emph{c)} for every vertex $v\in R_i$, the labels
match those of the corresponding vertices in $q$, $l(v) = l(f(v))$.  A query
workload is simply a multiset of these queries $Q = \{(q_1, n_1)\ldots (q_h,
n_h)\}$, where $n_i$ is the relative frequency of $q_i$ in $Q$.

A \textbf{query motif} is a graph which occurs, with a frequency
of more than some user defined threshold $\mathcal{T}$, as a sub-graph
of query graphs from a workload $Q$.

A vertex centric \textbf{graph partitioning} is defined as a disjoint family of sets
of vertices $P_k(G) = \{V_1,V_2,\ldots,V_k\}$. Each set $V_i$, together
with its edges $E_i$ (where $e_i \in E_i$, $e_i=(v_i, v_j)$, and
$\{v_i, v_j\} \subseteq V_i$), is referred to as a \emph{partition}
$S_i$. A partition forms a proper sub-graph of $G$ such that
$S_i = (V_i, E_i) $, $V_i \subseteq V$ and $E_i \subseteq E$.
We define the quality of a graph partitioning relative to a given workload $Q$.
Specifically, the number of inter-partition traversals ($ipt$) which occur while executing $Q$ over $P_k(G)$.
Whilst min. edge-cut is the standard scale free measure of partition
quality~\cite{Karypis1998a}, it is intuitively a proxy for $ipt$ and, as we have
argued (Sec.~\ref{section:introduction}), not always an effective one.

\subsection{Overview}
%
%Once again, we propose Loom, a system for iteratively producing a \emph{k}-way
%partitioning $P_k(G, Q)$ of an online graph $G$, given a workload $Q$, such that the \emph{probability} of \emph{ipt} when executing a random query $q\in Q$ is minimised.

Once again, Loom continuously partitions an online graph $G$ into $k$ parts,
optimising for a given workload $Q$.
The resulting partitioning $P_k(G, Q)$ reduces the \textit{probability} of
expensive $ipt$, when executing a random $q \in Q$, using the following
techniques.
%

%Note that $Q$ is a multiset, i.e.~may contain a distinct query multiple
%times; we denote this as $Q = \{(q_1, n_1)\ldots (q_h, n_h)\}$, where
%$n_i$ is the relative frequency of $q_i$ in $Q$.
%
%The remainder of this paper is organized as follows:
%

%
Firstly, we employ a trie-like datastructure to index all of the possible
sub-graphs of query graphs $q\in Q$, then identify those sub-graphs which are
motifs, i.e. occur most frequently
(Sec.~\ref{section:motif_mine}).
%
%First we employ a form of \emph{frequent sub-graph mining}
%\cite{Jiang2004a} using an efficient form of probabilistic isomorphism
%checking to build a trie-like datastructure, enumerating all possible
%sub-graphs of the query graphs $q\in Q$ then identifying the most
%frequent, which we call motifs $M$.
%
Secondly, we buffer a sliding window over $G$, then use an efficient graph stream pattern matching procedure to check whether each new edge added to $G$ creates a sub-graph which matches one of our motifs (Sec.~\ref{section:motif_match}).
Finally, we employ a combination of novel and existing partitioning heuristics
to assign each motif matching sub-graph which leaves the sliding window entirely
within an individual partition, thereby reducing $ipt$ for
$Q$ (Sec.~\ref{section:motif_alloc}).

\section{Identifying Motifs} \label{section:motif_mine}
\begin{figure}
	\centering
	\includegraphics[width=0.95\columnwidth]{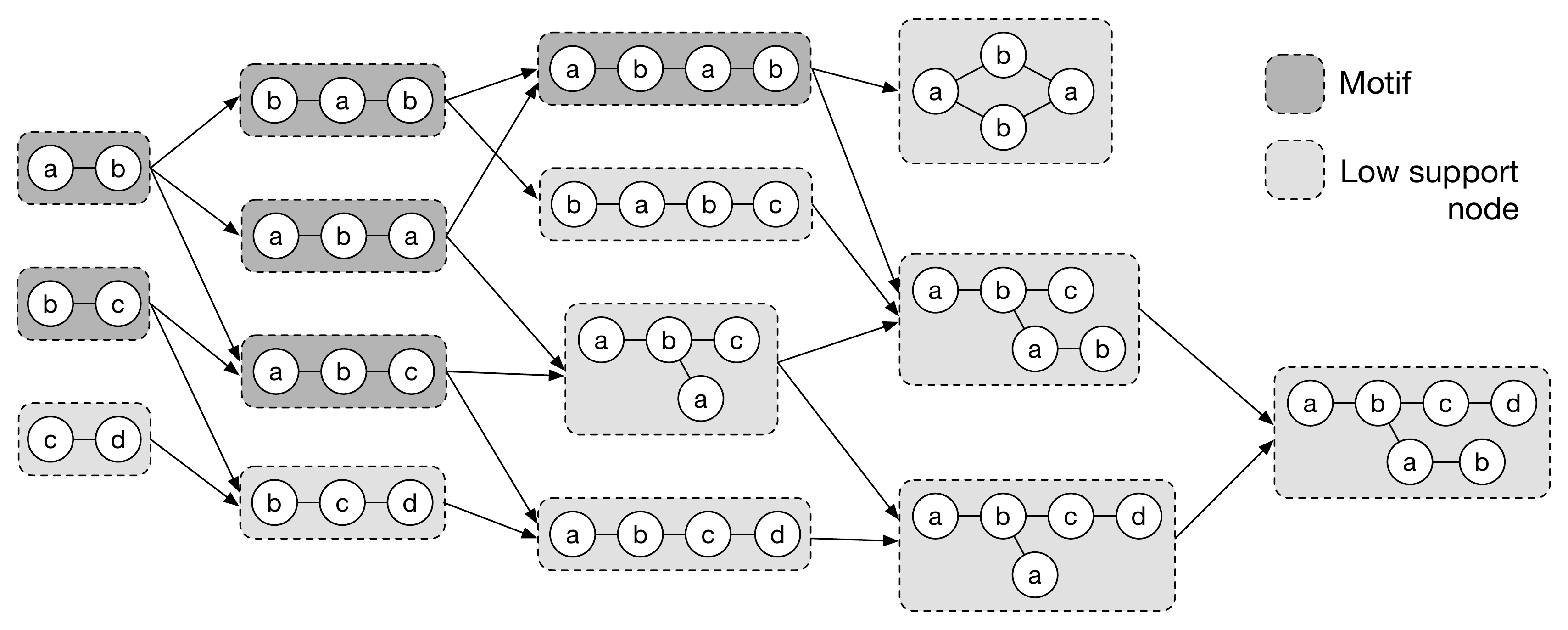}
	\caption{TPSTry++ for \Q{} in fig.~\ref{fig:graph_plus_workload}}
	\label{fig:tpstry_pp}	
\end{figure}

%\section{Identifying Motifs: capturing common sub-graphs in $Q$} \label{section:motif_mine}
%Lidl1997a
We now describe the first of the three steps mentioned above, namely the encoding of all query graphs found in our pattern matching query workload $Q$.
%
%\authornote{PM}{I assume you now introduced\\
%1- index Q and build trie, 2-detect motifs in the stream, and 3-allocate motifs to partitions \\if you have done this then there is no need to repeat the motivation.}
%
For this, we use a trie-like datastructure which we have called the \emph{Traversal Pattern Summary Trie} (TPSTry++).
%
%
%In order to detect the sub-graphs from a graph stream $G$ which are likely to be frequently traversed by our workload stream $Q$, we must first discover those sub-graphs which are common to many of the query graphs $q\in Q$. 
%%
%We achieve this by extrapolating all possible sub-graphs of each new query as it is executed, then encoding these in a trie-like datastructure which we have called the \emph{Traversal Pattern Summary Trie} (TPSTry++).
%
In a TPSTry++, every node represents a graph, while every parent node represents a sub-graph which is common to the graphs represented by its children.
%
%Conversely, each pattern in a child node contains one node more than each pattern in the parent nodes.
%
As an illustration, the complete TPSTry++ for the workload $Q$ in Fig.~\ref{fig:graph_plus_workload} is shown in Fig.~\ref{fig:tpstry_pp}.
%
%\authornote{PM}{``these have been ommitted from fig.~\ref{fig:tpstry_pp} for clarity.'' \\ why? I thought it'd be useful and you were going to add that.}

This structure not only encodes all sub-graphs found in each $q \in Q$, it also associates a support value $p$ with each of its nodes, to keep track of the relative frequency of occurrences of each sub-graph in our query graphs.

Given a threshold $\mathcal{T}$ for the frequency of occurrences, a \emph{motif}
is a sub-graph that occurs at least $\mathcal{T}$ times in $Q$.
As an example, for $\mathcal{T} = 40\%$, $Q$'s motifs are the shaded nodes in Fig.~\ref{fig:tpstry_pp}. 

%TODO: Fix the below - don't like it.
Intuitively, a sub-graph of $G$ which is frequently traversed by a query workload should be assigned to a single partition.  
We can idenfity these sub-graphs as they form within the stream of graph updates, by matching them against the motifs in the TPSTry++.
Details of the motif matching process are provided in Sec.\ref{section:motif_match}.
In the rest of this section we explain how a TPSTry++ is constructed, given a workload $Q$.

%
%
%%We associate a support value ($p$) with each node in the trie, though these have been ommitted from fig.~\ref{fig:tpstry_pp} for clarity.
%%
%When extrapolating the sub-graphs of each $q$, if a sub-graph has been seen before, and therefore already exists in the trie, we increment its suport value.
%%
%By specifying a minimum support value ${\cal T}$, we can filter the nodes of the trie and derive a set of all the sub-graphs which occur most often in $Q$, i.e. our \emph{motifs}.
%%
%Given an example ${\cal T}$ of $40\%$, $Q$'s motifs are presented as shaded nodes in fig.~\ref{fig:tpstry_pp}. 
%%
%The sub-graphs in $G$ which are isomorphic to these motifs are those most likely to be frequently traversed when executing queries from $Q$.
%%
%In other words, the filtered trie constitutes an intensional representation of the (likely large) set of ``popular'' sub-graphs in $G$.
%
%%

A TPSTry++ extends a simpler structure, called TPSTry, which we have recently
proposed in a similar setting \cite{Firth2017}.
It employs \emph{frequent sub-graph mining}\cite{Jiang2004a}  to compactly encode general labelled graphs.
The resulting structure is a Directed Acyclic Graph (DAG), to reflect the multiple ways in which a particular query pattern may extend shorter patterns. 
For example in Fig.~\ref{fig:tpstry_pp} the graph in node $a$-$b$-$a$-$b$ can be produced in two ways, by adding a single $a$-$b$ edge to either of the sub-graphs $b$-$a$-$b$, and $a$-$b$-$a$.
In contrast, a TPSTry is a tree that encodes the space of possible traversal \textbf{paths} through a graph as a conventional trie of strings, where a path is a string of vertex labels, and possible paths are described by a stream of regular path queries \cite{mendelzon1995finding}.

%
%Populating the TPSTry++ datastructure constitutes a form of \emph{frequent sub-graph mining} and extends both the work in that area\cite{?} and our own previous efforts\cite{?}\footnote{Paper submitted for publication elsewhere}, in which we define the original TPSTry.
%
%The original TPSTry datastructure compactly encodes the space of possible traversal \textbf{paths} through a graph as a conventional trie of strings, where a path is a string of vertex labels, and possible paths are described by a stream of regular path queries \cite{?}.
%%
%The TPSTry++ is similar, however it encodes \textbf{general labelled graphs} (not just paths), which are inherently unordered. 
%
%For example, consider the graph $q_3$ from fig.~\ref{fig:graph_plus_workload}: it has no obvious first element.
%%
%Furthermore, for a path of length $n$ any prefixes of length $n-1$ are guaranteed to be equal, whereas for a graph with $n$ edges fig.~\ref{fig:tpstry_pp} demonstrates that sub-graphs of size $n-1$ may be distinct.
%%
%This means that for each query graph encoded in the TPSTry++, there are potentially many paths from the root node, representing many possible orders of construction, and makes the datastructure a \emph{directed acyclic graph} (DAG) rather than a tree.
%

%Note that the TPSTry++ may be viewed as a collection of overlapping lattices, one for each query graph, defined by the \emph{is sub-graph of} relationship; i.e. every parent node in the the trie represents a sub-graph of each of its children.
%%
\begin{figure}
	\includegraphics[width=0.9\columnwidth]{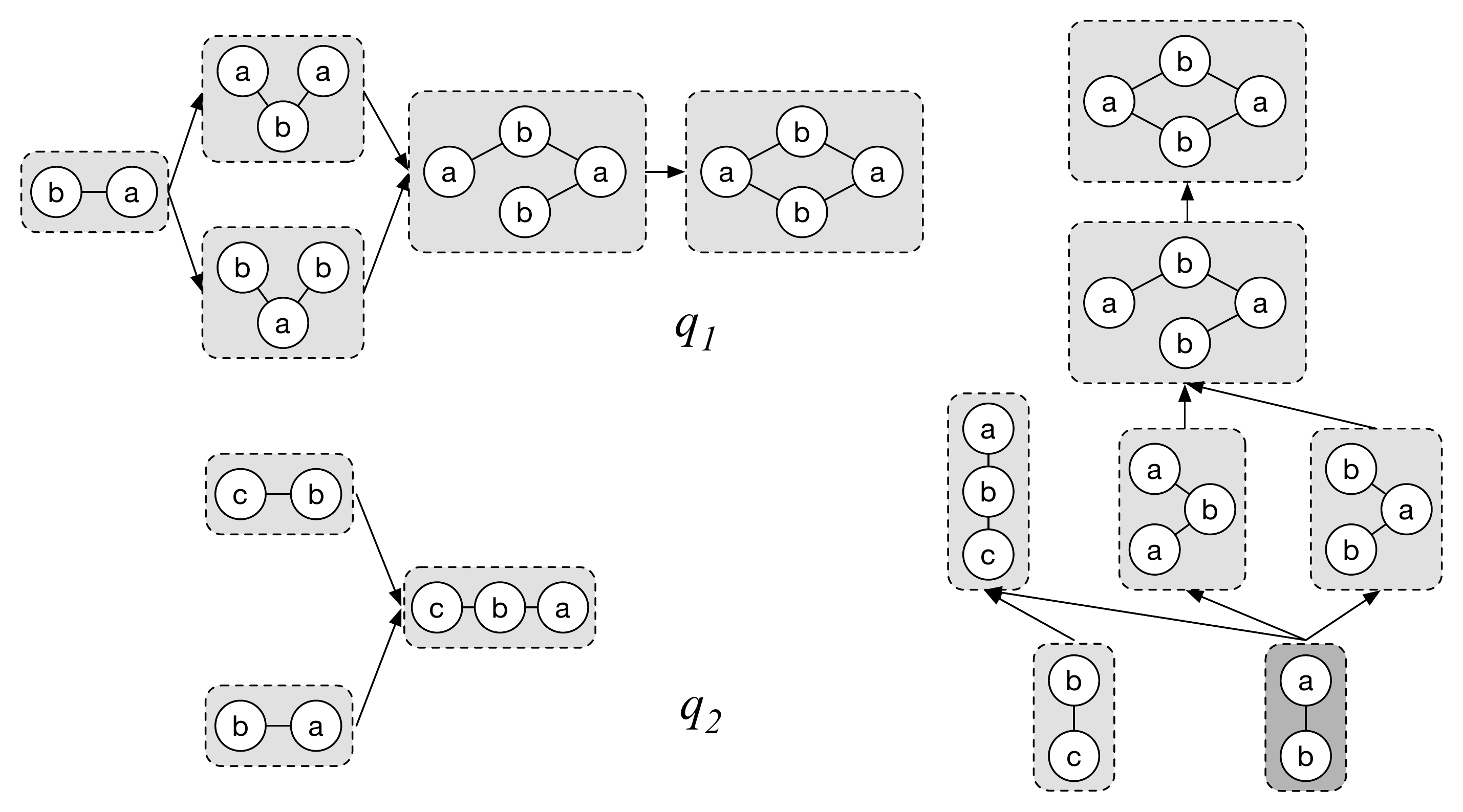}
	\caption{Combining tries for query graphs $q_1$, $q_2$}
	\label{fig:q_lattice_merging}
\end{figure}
Note that the trie is a relatively compact  structure, as it grows with $|L_V|^t$, where $t$ is the number of edges in the largest query graph in $Q$ and $L_V$ is typically small.
Also note that the TPSTry++ is similar to, though more general than, Ribiero et al's G-Trie~\cite{Ribeiro2010} and Choudhury et al's SJ-Tree~\cite{Choudhury2015a}, which use trees (not DAGs) to encode unlabelled graphs and labelled paths respectively.
\subsection{Sub-graph signatures}~\label{subsection:sub_graph_signatures}
We build the trie for $Q$ by progressively building and merging smaller tries
for each $q \in Q$, as shown in Fig.~\ref{fig:q_lattice_merging}.
%\authornote{PM}{PLEASE CHECK this is correct??}
%presents an example of how we construct a smaller trie by merging to sub-graph lattices of two query graphs. 
%
%
This process relies on detecting graph isomorphisms, as any two trie nodes from different queries that contain identical graphs should be merged.
Failing to detect isomorphism would result, for instance, in two separate trie nodes being created for the simple graphs $a$-$b$-$c$ and $c$-$b$-$a$, rather than a single node with a support of 2, as intended.
One way of detecting isomorphism, often employed in frequent sub-graph mining, involves computing the lexicographical canonical form for each graph \cite{McKay1981a}, whereby two graphs are isomorphic if and only if they have the same canonical representation.

%when adding a new query graph $q_i$, if nodes which represents $q_i$'s sub-graphs already exist elsewhere in the trie, then we should merge them.
%
%In order to construct the TPSTry++, we need to employ some form of pattern matching to detect if two sub-graphs are isomorphic.
%
%Using this isomorphism checking, when adding a new query graph $q_i$, if nodes which represents $q_i$'s sub-graphs already exist elsewhere in the trie, then we should merge them.
%
%Otherwise, for instance, we would create two separate trie nodes for the simple graphs $a$-$b$-$c$ and $c$-$b$-$a$, rather than a single node with a support of 2, as intended.
%
%Frequent sub-graph mining techniques often compute each graph's lexicographical canonical form for this purpose\cite{?}, where the canonical form of two isomorphic graphs is guaranteed to be equal and that of two non-isomorphic graphs is guaranteed \textbf{not} to be. 
%
%
Computing a graph's canonical form provides strong guarantees, but  can be expensive\cite{Ribeiro2010}. 
Instead, we propose a probabilistic, but computationally more efficient approach based on \emph{number theoretic signatures}, which extends recent work by Song et al.~\cite{Song2014}.
In this approach we compute the signature of a graph as a large, pseudo-unique integer hash that encodes key information such as its vertices, labels, and nodes degree. 
Graphs with matching signatures are likely to be isomorphic to one another, but there is a small probability of collision, i.e., of having two different graphs with the same signature.
%
% 
%as an alternative, we propse an efficient \emph{number theoretic signatures} approach, extending the recent work of Song et al.~\cite{?}.
%%
%Essentially we compute a large, pseudo-unique integer hash for a given graph, capturing key information such as vertices, their labels and their degree, as distinct factors. 
%%
%If two graphs' signatures match, they are \emph{likely} to be isomorphic to one another, though this is not guaranteed and therefore signatures do not constitute a canonical form.
%%

%
Given a query graph $G_q = \{V_q, E_q \}$ we compute its signature as follows.
Initially we assign a random value $r(l) = [1, p)$, between 1 and some user specified prime $p$, to each possible label $l\in L_{V_i}$ from our data graph $G$; recall that the function $f_\vertexlabel$ maps vertices in $G$ to these labels.
We then perform three steps: 
\begin{enumerate}
    \item Calculate a factor for each edge $e = (v_i, v_j)\in E_q$, according to the formula: 
	    \[ edgeFac(e) = (r(f_\vertexlabel{}(v_i)) - r(f_\vertexlabel{}(v_j)))\  mod\ p \]
    \item Calculate the factors that encode the degree of each vertex. If a
	    vertex $v$ has a degree $n$, its degree factor is defined as: 
	    \begin{align*}
		    & degFac(v) = ((r(f_\vertexlabel{}(v)) + 1)\ mod\ p)\cdot \\ 
			((r(f_\vertexlabel{}(v & )) + 2)\ mod\ p)\cdot \ldots \cdot ((r(f_\vertexlabel{}(v)) + n)\ mod\ p) 
	    \end{align*}	    
    \item Finally, we compute the signature of $G_q = (V_q, E_q)$ as: $$(\prod_{e\in E_i} edgeFac(e))\cdot{}(\prod_{v\in V_i} degFac(v))$$
\end{enumerate}

To illustrate, consider query $q_1$ from Fig.~\ref{fig:graph_plus_workload}.
Given a $p$ of 11 and random values $r(a) = 3$, $r(b) = 10$ we first calculate the edge factor for an $a$-$b$ edge: $edgeFac((a,b)) = (3 - 10)\ mod\ 11 = 7$. 
As $q_1$ consists of four $a$-$b$ edges, its total edge factor is $7^4 = 2401$.
Then we calculate the degree factors~\footnote{Note we don't consider 0 a valid
factor, and replace it with $p$ (e.g. $11\ mod\ 11 = 11$)}, starting with a $b$ labelled vertex with
degree 2: $degFac(b) = ((10 + 1)\ mod\ 11)\cdot ((10 + 2)\ mod\ 11) = 11$,
followed by an $a$ labelled vertex also with degree 2: $degFac(a) = 20$.
%
%As there are two $a$ and two $b$ vertices, all with degree 2, the total degree factor is .
As there are two of each vertex, with the same degree, the total degree factor
is $11^2 \cdot 20^2 = 48400$.
The signature of $q_1 = 2401 \cdot 48400 = 116208400$.
This approach is appealing for two reasons.
Firstly, since the factors in the signature may be multiplied in any order, a signature for $G$ can be calculated incrementally if the signature of any of its sub-graphs $G_i$ is known, as this is the combined factor due to the additional edges and degree in $G \setminus G_i$. 
%
%Song et al. use this property of signatures to perform graph stream pattern matching\cite{?}, iteratively computing the signature of any connected sub-graph in a stream window as new edges are added. 
%%
%They then check if the signature is divisible by the signature of a known query graph; if so, then the sub-graph should contain a query match.  
%
Secondly, the choice of $p$ determines a trade-off between the probability of collisions and the performance of computing signatures. 
Specifically, note that signatures can be \textbf{very} large numbers (thousands of bits) even for small graphs, rendering operations such as remainder costly and slow.
A small choice of $p$ reduces signature size, because all the factors are mapped to a finite field~\cite{Lidl1997a} (\emph{factor mod p}) between 1 and $p$, but it increases the likelihood of collision, i.e., the probability of two unrelated factors being equal.
We discuss how to improve the performance and accuracy of signatures in Section~\ref{subsection:avoiding_signature_collisions}.
\subsection{Constructing the TPSTry++}
\begin{algorithm}
\small
\caption{Recursively add a query graph $G_q$ to a TPSTry++}
\label{algo:weave}
\begin{algorithmic}[1]
\STATE $factors(e, g) \gets$ degree/edge factors to multiply a graph $g$'s signature when adding edge $e$
\STATE $support(g) \gets$ a map of TPSTry++ nodes (graphs) to $p$-values
	\STATE $tpstry \gets$ the TPSTry++ for workload \Q{}
\STATE $parent \gets$ a TPSTry++ node, initially root (an empty graph)
\STATE $G_q \gets$ the query graph defined by a query $q$
\STATE $g$ some sub-graph of $G_q$ 
\\ \hfill
%\STATE $\mathbf{weave(G_q, tpstry)}$
    \STATE $\mathbf{for}$ $e$ in edges from $G_q$ $\mathbf{do}$
        \INDSTATE $\mathbf{g} \gets$ new empty graph
        \INDSTATE $\mathbf{corecurse(parent, e, tpstry, g)}$
            \INDSTATE[2] $sig \gets factor(e, g)\cdot parent.signature$
            \INDSTATE[2] $\mathbf{if}$ $tpstry.signatures$ contains $sig$ $\mathbf{then}$ 
            	\INDSTATE[3] $n \gets$ node from $tpstry$ with signature $sig$
                \INDSTATE[3] $n.support \gets$ $n.support + 1$
            \INDSTATE[2] $\mathbf{else}$ 
                \INDSTATE[3] $n \gets$ new node with graph $g+e$, signature $sig$ and support 1
                \INDSTATE[3] $tpstry \gets tpstry + n$
            \INDSTATE[2] $\mathbf{if~not}$ $parent.children$ contains $n$ $\mathbf{then}$
                \INDSTATE[3] $parent.children \gets$ $parent.children + n$
            \INDSTATE[2] $newEdges \gets$ edges incident to $g+e$ \& not in $g+e$
            \INDSTATE[2] $\mathbf{for}$ $e'$ in $newEdges$
                \INDSTATE[3] $\mathbf{corecurse(n, e', tpstry, g+e)}$
    \STATE $\mathbf{return}$ $tpstry$  
\end{algorithmic}
\end{algorithm}
Our approach to constructing the TPSTry++ is to incrementally compute signatures for sub-graphs of each query graph $q$ in a trie, merging trie nodes with equal signatures to produce a DAG which encodes the sub-graphs of all $q\in Q$. 
Alg.~\ref{algo:weave} formalises this approach.
%
%
%As a result of the iterative nature of signature computation, given each new query graph $G_q = \{V_q, E_q\} $ from a workload stream $Q$, we are able to construct a signature for each of its sub-graphs then add them all to TPSTry++. 
%%
%Alg.~\ref{algo:weave} presents the simple corecursive procedure for constructing or adding to the trie, which we refer to as \emph{weave}.
%

Essentially, we recursively ``rebuild'' the graph $G_q$ $\mid Eq \mid$ times, starting from each edge $e\in E_q$ in turn.
For an edge $e$ we calculate its edge and degree factors, initially assuming a degree of 1 for each vertex.
If the resulting signature is not associated with a child of the TPSTry++'s root, then we add a node $n$ representing $e$.
Subsequently, we ``add'' those edges $e'$ which are incident to $e \in G_q$, calculating the additional edge and degree factors, and add corresponding trie nodes as children of $n$.
Then we recurse on the edges incident $e+e'$.
%

% 
%\textit{Example.} % Could use the Example pattern, but haven't throughout the rest of the paper, so for now lets go with consistency.
Consider again our earlier example of the query graph $q_1$: as it arrives in the workload stream $Q$, we break it down to its constituent edges \emph{\{a-b, a-b, a-b, a-b\}}. 
Choosing an edge at random we calculate its combined factor.
We know that the edge factor of an $a$-$b$ edge is 7.
When considering this single edge, both $a$ and $b$ vertices have a degree of 1, therefore the signature  for $a$-$b$ is $7\cdot ((3 + 1)\ mod\ 11)\cdot ((10 + 1)\ mod\ 11) = 308$.
Subsquently, we do the same for all other edges and, finding that they have the same signature, leave the trie unmodified. 
%increment the $a$-$b$ node's support in the trie.
%
Next, for each edge, we add each incident edge from $q_1$ and compute the new combined signature.
Assume we add another $a$-$b$ edge adjacent to $b$ to produce the sub-graph $a$-$b$-$a$.
This produces three new factors: the new edge factor $7$, the new $a$ vertex degree factor $((3 + 1)\ mod\ 11)$ and an additional degree factor for the existing $b$ vertex $((10 + 2)\ mod\ 11)$. 
The combined signature for $a$-$b$-$a$ is therefore $308\cdot 7\cdot 4\cdot 1 = 8624$; if a node with this signature does not exist in the trie as a child of the $a$-$b$ node, we add it.
This continues recursively, considering larger sub-graphs of $q_1$ until there are no edges left in $q_1$ which are not in the sub-graph, at which point, $q_1$ has been added to the TPSTry++.
%

%Remember that Loom's graph partitioning is intended to work in an online
%environment, where new edges continuously arrive. We aim to place these
%within partitions which will minimise the number of expensive
%inter-partition traversals when executing the current workload $Q$. In
%such an online setting, the characteristics of $Q$ are similarly liable
%to change. We must continually update the structure and support values
%of the TPSTry++ to reflect these changes, in order to avoid Loom making
%poor edge placement decisions .
%
%We associate each new query with a unique key, and maintain a table
%mapping queries to their respective frequencies. These frequencies are
%approximated using a sketch datastructure which samples the occurrences
%of each query within a sliding window of time \emph{t} over the stream
%$Q$. In order to ensure that TPSTry++ nodes have up-to-date support
%values, we do not store support values in the nodes directly. Instead
%each node \emph{n} is associated with the set of keys which correspond
%to queries which include \emph{n}'s sub-graph.
%
\subsection{Avoiding signature collisions} \label{subsection:avoiding_signature_collisions}
As mentioned, number theoretic signatures are a probabilistic method of ismorphism checking, prone to collisions.
There are several scenarios in which two non-isomorphic graphs may have the same signature: \emph{a)} two factors representing different graph features, such as different edges or vertex degrees, are equal; \emph{b)} two distinct sets of factors have the same product; and \emph{c)} two different graphs have identical sets of edges, vertices and vertex degrees.
%Permutations of a multiset?

The original approach to graph isomorphic checking \cite{Song2014} makes use of
an expensive authoritative pattern matching method to verify identified matches. 
Given a query graph, it calculates its signature in advance, then incrementally computes signatures for sub-graphs which form within a window over a graph stream.
If a sub-graph's signature is ever divisible by that of the query graph, then that sub-graph should contain a query match. 

There are some key differences in how we compute and use signatures with Loom, which allow us to rely solely upon signatures as an efficient means for mining and matching motifs.
Firstly, remember our overall aim is to heuristically lower the probability that sub-graphs in a graph $G$ which match our discovered motifs straddle a partition boundary.
As a result we can tolerate some small probability of false positive results, whilst the manner in which signatures are executed (Sec.~\ref{subsection:sub_graph_signatures}) precludes false negatives; i.e. two graphs which \textbf{are} isomorphic are guaranteed to have the same signature. 
Secondly, we can exploit the structure of the TPSTry++ to avoid ever explicitly computing graph signatures.
From Fig.~\ref{fig:tpstry_pp} and Alg.~\ref{algo:weave}, we can see that all possible sub-graphs of a query graph $G_q$ will exist in the TPSTry++ by construction.
We calculate the edge and degree factors which would multiply the signature of a sub-graph with the addition of each edge, then associate these factors to the relevant trie branches.
This allows us to represent signatures as sets of their constituent factors, which eliminates a source of collisions, e.g. we can now distinguish between graphs with factors $\{6,2\}$, $\{4,3\}$ and $\{12\}$.
Thirdly, we never attempt to discover whether some sub-graph \textbf{contains} a match for query $q$, only whether it \textbf{is} a match for $q$.
In other words, the largest graph for which we calculate a signature is the size of the largest query graph $\vert G_q \vert$ for all $q\in Q$, which is typically small\footnote{Of the order of 10 edges.}.
This allows us to choose a larger prime $p$ than Song et al. might, as we are less concerned with signature size, reducing the probability of factor collision, another source of false positive signature matches.
Concretely, we wish to select a value of $p$ which minimises the probability
that more than some acceptable percentage $\mathcal{C}\%$ of a signature's factors are collisions.
From Section~\ref{subsection:sub_graph_signatures} there are three scenarios in which a factor collision may occur: \emph{a)} two edge factors are equal despite different vertices with different random values from our range $[1, p)$; \emph{b)} an edge factor is equal to a degree factor; and \emph{c)} two degree factors are equal, again despite different vertices. 
Song et al. show that all factors are uniform random variables from $[1, p)$,
therefore each scenario occurs with probability $\frac{1}{p}$.

For either edge or degree factors, from the above it is clear that there are two scenarios in which a collision may occur, giving a collision probability for any given factor of $\frac{2}{p}$.
The Handshaking lemma tells us that the total degree of a graph must equal $2\vert E\vert$, which means that a graph must have $3\vert E\vert$ factors in its signature: one per edge plus one per degree.
Combined with the binary measure of ``success'' (collision / no collision), this suggests a binomial distribution of factor collision probabilities, specifically $Binomial(3\vert E\vert, \frac{2}{p})$.
Binomial distributions tell us the probability of exactly $x$ ``successes''
occuring, however we want the probability that no more than $\mathcal{C}_{max} =
\mathcal{C}\% \cdot 3\vert E\vert$ factors collide and so must sum over all
acceptable outcomes $x\in \mathcal{C}_{max}$:
\[ 
	\sum_{x = 0}^{\mathcal{C}_{max}} Pr(X = x) \text{ where } X\sim Binomial(3\vert E\vert ,\frac{2}{p})
\]
\begin{figure}
	\includegraphics[width=0.90\columnwidth]{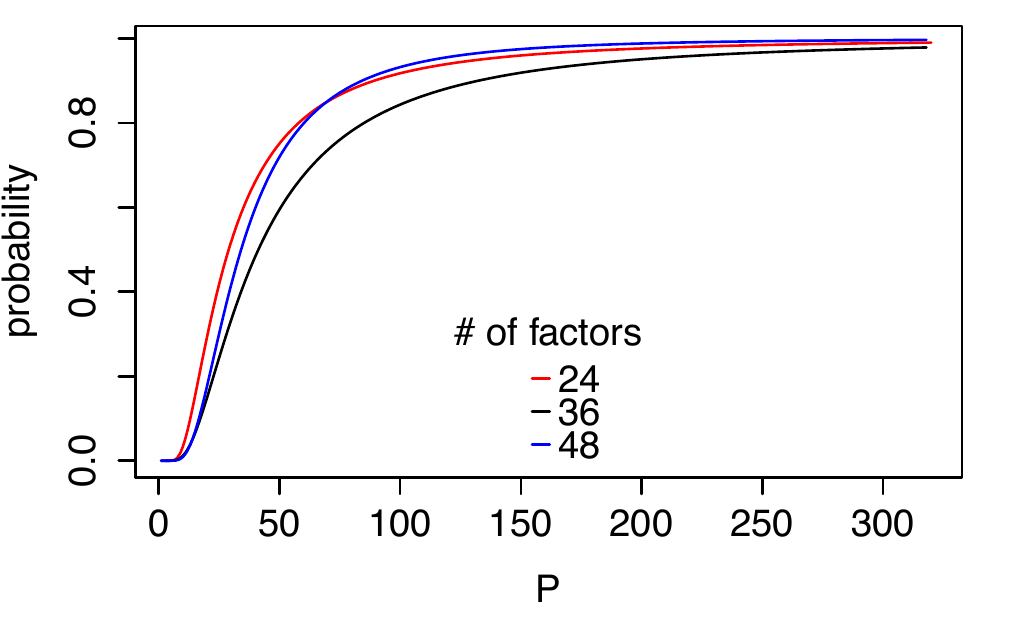}
	\caption{Probability of $<5\%$ factor collisions for various $p$}
	\label{fig:p_choices}
\end{figure}
Figure~\ref{fig:p_choices} shows the probabilities of having fewer than $5\%$ factor collisions given query graphs of 8, 12 or 16 edges and $p$ choices between 2 and 317.
In Loom, when identifying and matching motifs, we use a $p$ value of 251, which as you can see gives a neglible probability of significant factor collisions.

\section{Matching Motifs} \label{section:motif_match}
%\section{Matching Motifs: finding hot sub-graphs in a graph stream}  \label{section:motif_match}
%
%As in fig.~\ref{fig:tpstry_pp}, once we have constructed the TPSTry++ we can filter those nodes with a support $p$-value less than some threshold $\mathcal{T}$ in order to discover our motifs given the query stream $Q$.
%
%Recall that, with Loom's partitioning, we aim to optimise the performance of sub-graph pattern matching queries by reducing the number of expensive inter-partition traversals.
%
We have seen how motifs that occur in $Q$ are identified.
By construction, motifs represent graph patterns that are frequently traversed during executions of queries in $Q$.
Thus, the sub-graphs of $G$ that match those motifs are expected to be frequently visited together and are therefore best placed within the same partition. 
In this section we clarify how we discover pattern matches between sub-graphs and motifs, whilst in the next Section we describe the allocation of those sub-graphs to partitions.
%

%Many approaches exists for matching sub-graphs, often following the \emph{filter} then \emph{verify} pattern \cite{?}. 
%
%Verifying matches normally involves traversing the edges of candidate sub-graphs. 
%
%Practically, this is the reason sub-graphs in a graph $G$ which are isomorphic to motifs are likely to be frequently traversed during execution of any $q\in Q$.
%
%
Loom operates on a sliding window of configurable size over the stream of edges that make up the growing graph $G$.
The system monitors the connected sub-graphs that form in the stream within the space of the window, efficiently checking for isomorphisms with any known motif each time a sub-graph grows.
Upon leaving the window, sub-graphs that match a motif are immediately assigned to a partition, subject to partition balance constraints as explained in Section~\ref{section:motif_alloc}.

Note that this technique introduces a delay, corresponding to the size of the window, between the time edges are submitted to the system and the time they are assigned and made available. 
In order to allow queries to access the new parts of graph $G$, Loom views the sliding window \textit{itself} as an extra partition, which we denote $P_{temp}$.
In practice, vertices and edges in the window are accessible in this temporary partition prior to being permanently allocated to their own partition.
% Say something about width of sliding window finding more matches at the cost of memory. "Also note ... intu big t == better ..."

%
To help understand how the matching occurs, note that in the TPSTry++, by construction, \textit{all} anscestors of any node $n$ \textit{must} represent strict sub-graphs of the graph represented by $n$ itself.
Also, note that the support of a node $n$ is the relative frequency with which $n$'s sub-graph $G_n$ occurs in $Q$. 
As, by definition, each time $G_n$ occurs in $Q$ so do all of \textit{its} sub-graphs, a trie node $n$ must have a support lower than any of its anscestors.
This means that if any of the nodes in the trie, including those representing single edges, are not motifs, then none of their descendants can be motifs either. 
Thus, when a new edge $e=(v_1, v_2)$ arrives in the graph stream, we compute its signature (Sec.~\ref{subsection:sub_graph_signatures}) and check if $e$ matches a single-edge motif at the root of the TPSTry++. 
If there is no match, we can be certain that $e$ will never form part of any sub-graph that matches a motif.
We therefore immediately assign $e$ to a partition and do not add it to our stream window $P_{temp}$.
If, on the other hand, $e$ does match a single-edge motif then we record the match into a map, \emph{matchList}, and add $e$ to the window.
The \emph{matchList} maps vertices $v$ to the set of motif matching sub-graphs in $P_{temp}$ which contain $v$; i.e. having determined that $e = (v_1, v_2)$ is a motif match, we treat $e$ as a sub-graph of a single edge, then add it to the \emph{matchList} entries for both $v_1$ and $v_2$.
Additionally, alongside every sub-graph in \emph{matchList}, we store a reference to the TPSTry++ node which represents the matching motif. 
Therefore, entries in \emph{matchList} take the form $v\rightarrow \{\langle E_i, m_i\rangle , \langle E_j, m_j\rangle , \ldots \}$, where $E_i$ is a set of edges in $P_{temp}$ that form a sub-graph $g_i$ with the same signature as the motif $m_i$.
% Mention why we index by vertices as the perfect lead in to wanting to discover larger motifs

%
\begin{figure*}
\centering
	\includegraphics[width=2\columnwidth]{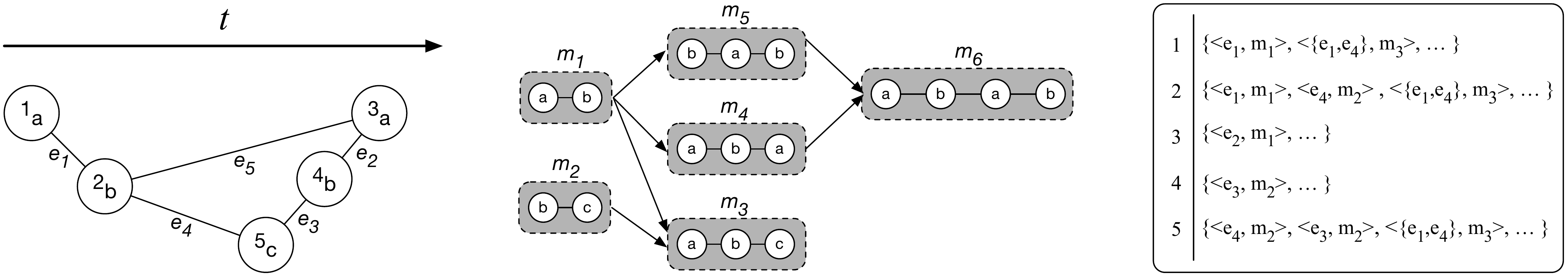}
	\caption{$t$-length window over $G$ (left), Motifs from TPSTry++ (center) and motif $matchList$ for window (right)}
	\label{fig:matchList}
\end{figure*}
Given the above, any edge $e$ which is added to $P_{temp}$ must at least match a single edge motif.
However, if $e$ is incident to other edges already in $P_{temp}$, then its addition may also form larger motif matching sub-graphs which we must also detect and add to \emph{matchList}.
Thus, having added $e = (v_1, v_2)$ to \emph{matchList}, we check the map for existing matches which are connected to $e$; i.e we look for matches which contain one of $v_1$ or $v_2$.
If any exist, we use the procedure in Alg.~\ref{algo:mine_motif_matches}, along with the TPSTry++, to determine whether the addition of edge $e$ to these sub-graphs creates another motif match.
Essentially, for each sub-graph $g_i$ from \emph{matchList} to which $e$ is connected, we calculate the set of edge and degree factors $fac(e, g_i)$ which would multiply the signature of $g_i$ upon the addition $e$, as in Sec.~\ref{section:motif_mine}.
Recall, also from Sec.~\ref{section:motif_mine}, that a TPSTry++ node contains a signature for the graph it represents, and that these signatures are stored as sets of factors, rather than their large integer products.
As each sub-graph in \emph{matchList} is paired with its associated motif $n$ from the trie, we can efficiently check if $n$ has a child $c$ where \emph{a)} $c$ is a motif; and \emph{b)} the difference between $n$'s factor set and $c$'s factor set corresponds to factors for the addition of $e$ to $g_i$, i.e., $fac(e, g_i) = c.signatures\setminus n.signatures$.
\begin{algorithm}[t]
\small
\caption{Mine motif matches from each new edge $e\in G$}
\label{algo:mine_motif_matches}
\begin{algorithmic}[1]
\STATE $fac(e, g) \gets$ degree/edge factors to multiply a graph $g$'s signature when adding edge $e$
\STATE $tpstry \gets$ the filtered TPSTry++ of motifs for workload $Q$
\\ \hfill	
\STATE $\mathbf{for}$ each new edge $e(v_1, v_2)$ $\mathbf{do}$
\INDSTATE[0.8] $matches \gets matchList(v_1) \cup matchList(v_2)$
\INDSTATE[0.8] $\mathbf{for}$ each sub-graph $m$ in $matches$ $\mathbf{do}$ 
\INDSTATE[1.6] $n \gets $ the $tpstry$ node for $m$
\INDSTATE[1.6] $\mathbf{if}$ $n$ has child $c$ w. $factor = fac(e, m)$ $\mathbf{then}$
\INDSTATE[2.4] add $\langle m + e, c\rangle $ to $matchList$ for $v_1\& v_2$ //Match found!
\INDSTATE[0.8] $ms_1 \gets matchList(v_1)$
\INDSTATE[0.8] $ms_2 \gets matchList(v_2)$ 
\INDSTATE[0.8] $\mathbf{for}$ all possible pairs $(m_1, m_2)$ from $(ms_1, ms_2)$ $\mathbf{do}$
\INDSTATE[1.6] $n_1 \gets $ the $tpstry$ node for $m_1$ 
\INDSTATE[1.6] $\mathbf{recurse(tpstry, m_2, m_1, n_1)}$
\INDSTATE[2.4] $\mathbf{for}$ each edge $e_2$ in $m_2$ $\mathbf{do}$
\INDSTATE[3.2] $\mathbf{if}$ $n_1$ has child $c_1$ w. $factor = fac(e_2, m_1)$ $\mathbf{then}$
	\INDSTATE[4] $\mathbf{recurse(tpstry, m_2 - e_2, m_1 + e_2, c_1)}$ 
\INDSTATE[2.4] $\mathbf{if}$ $m_2$ is empty $\mathbf{then}$ //Match found!
	\INDSTATE[3.2] add $\langle m_1 + m_2, n_1\rangle $ to $matchList$ for $v_1\& v_2$ 
\end{algorithmic}
\end{algorithm}
If such a child exists in the trie then adding $e$ to a graph which matches motif $n$ ($g_i$) will likely create a graph which matches motif $c$: the addition of $e$ to $P_{temp}$ has formed the  new motif matching sub-graph $g_i + e$.
We also detect if the joining of two existing multi edge motif matches $(\langle E_1, m_1\rangle , \langle E_2, m_2\rangle )$ forms yet another motif match, in roughly the same manner.
First we consider each edge from the smaller motif match (e.g. $e\in E_2$ from $\langle E_2, m_2\rangle $), checking if the addition of any of these edges to $E_1$\footnote{Treating $E_1$ as a sub-graph.} constitutes yet another match; if it does then we add the edge to $E_1$ and recursively repeat the process until $E_2$ is empty.
If this process \textbf{does} exhaust $E_2$ then $E_1\cup E_2$ constitute a motif matching sub-graph.
Once this process is complete, \emph{matchList} will contain entries for \textbf{all} of the motif matching sub-graphs currently in $P_{temp}$. 
Note that as more edges are added to $P_{temp}$, \emph{matchList} may contain
multiple entries for a given vertex where one match is a sub-graph of another,
i.e. new motif matches don't replace existing ones.
As an example of the motif matching process, consider the portion of a graph stream (left), motifs (center) and \emph{matchList} (right) depicted in Fig.~\ref{fig:matchList}.
Our window over the graph stream $G$ is initially empty, with the depicted edges being added in label order (i.e. $e_1, e_2,\ldots$). 
As the edge $e_1$ is added, we first compute its signature and verify whether $e_1$ matches a single-edge motif in the TPSTry++.
We can see that, as an $a$-$b$ labelled edge, the signature for $e_1$ must match that of motif $m_1$, therefore we add $e_1$ to $P_{temp}$, and add the entry $\langle e_1, m_1\rangle $ to \emph{matchList} for both $e_1$'s vertices 1,2.  
As $e_1$ is not yet connected to any other edges in $P_{temp}$, we do not need to check for the formation of additional motif matches.
Subsequently, we perform the exact same process for edge $e_2$.
When $e_3$ is added, again we verify that, as a $b$-$c$ edge, $e_3$ is a match for the single-edge motif $m_3$ and so update $P_{temp}$ and \emph{matchList} accordingly.
However, $e_3$ is connected to existing motif matching sub-graphs in $P_{temp}$ therefore the union of \emph{matchList} entries for $e_3$'s vertices 4,5 (line 4 Alg.~\ref{algo:mine_motif_matches}) returns $\{ \langle e_2, m_1\rangle \}$.
As a result, we calculate the factors to multiply $e_2$'s signature by, when adding $e_3$. 
Remember that when computing signatures, each edge has a factor, as well as each degree. 
Thus, when adding $e_3$ to $e_2$ our \emph{new} factors are an edge factor for a $b$-$c$ labelled edge, a first degree factor for the vertex labelled $c$ (5) and a second degree factor for the vertex labelled $b$\footnote{As, with the addition of $e_3$, vertex 4 has degree 2.} (4) (Sec.~\ref{subsection:sub_graph_signatures}).
Subsquently we must check whether the motif for $e_2$, $m_1$, has any child nodes with additional factors consistent with the addition of a $b$-$c$ edge, which it does: $m_3$.
This means we have found a new sub-graph in $P_{temp}$ which matches the motif $m_3$, and must add $\langle \{ e_2, e_3 \} , m_3\rangle $ to the \emph{matchList} entries for vertices 3, 4 and 5.
Similarly, the addition of $b$-$c$ labelled edge $e_4$ to our graph stream produces the new motif matches $\langle e_4, m_2\rangle$ and $\langle \{ e_1, e_4\} , m_3\rangle  $, as can be seen in our example \emph{matchList}.
Finally, the addition of our last edge, $e_5$, creates several new motif matches (e.g. $\langle \{ e_1, e_5\} , m_4\rangle , \langle \{ e_2, e_5\} , m_5\rangle $ etc\ldots ).
In particular, notice that the addition of $e_5$ creates a match for the motif $m_6$, combining the new motif match $\langle \{ e_1, e_5\} , m_4\rangle $ with an existing one $\langle e_2, m_1\rangle $.
To understand how we discover these slightly more complex motif matches, consider Alg.~\ref{algo:mine_motif_matches} from line 11 onwards.
First we retrieve the updated \emph{matchList} entries for vertices 2 and 3, including the new motif matches gained by simply adding the single edge $e_5$ to connected existing motif matches, as above.
Next we iterate through all possible pairs of motif matches for both vertices.
Given the pair of matches $(\langle \{ e_1, e_5\} , m_4\rangle , \langle e_2, m_1\rangle )$, we discover that the addition of any edge from the smaller match (i.e. $e_2$) to the larger produces factors which correspond to a child of $m_4$ in the TPSTry++: $m_6$.
As $e_2$ is the only edge in the smaller match, we simply add the match $\langle \{ e_1, e_2, e_5 \} , m_6\rangle $ to the \emph{matchList} entries for 1, 2, 3 and 4. 
In the general case however, we would not add this new match but instead recursively ``grow'' it with new edges from the smaller match, updating \emph{matchList} \emph{only} if all edges from the smaller match have been successfully added.  

\section{Allocating Motifs} \label{section:motif_alloc}
%\section{Allocating Motifs: partitioning graph stream to minimise $ipt$} \label{section:motif_alloc}
%
Following graph stream pattern matching, we are left with a collection of sub-graphs, consisting solely of the most recent $t$ edges in $G$, which match motifs from $Q$. 
As new edges arrive in the graph stream, our window $P_{temp}$ grows to size $t$ and then ``slides'', i.e. each new edge added to a full window causes the oldest ($t+1^{th}$) edge $e$ to be dropped.
Our strategy with Loom is to then assign this old edge $e$ to a permanent partition, along with the other edges in the window which form motif matching sub-graphs with $e$.
The sole exception to this is when an edge arrives that may not form part of any motif match and is assigned to a partition immediately (Sec.~\ref{section:motif_match}). 
This exception does not pose a problem however, because Loom behaves as if the edge was never added to the window  and therefore does not cause displacement of older edges.
Recall again that with Loom we are attempting to assign motif matching sub-graphs wholly within individual partitions with the aim of reducing $ipt$ when executing our query workload $Q$.
One naive approach to achieving this goal is as follows: 
When assigning an edge $e = (v_1, v_2)$, retrieve the motif matches associated with $v_1$ and $v_2$ from $P_{temp}$ using our $matchList$ map, then select the subset $M_e$ \textbf{that contains} $e$, where $M_e = \{\langle E_1,m_1\rangle ,\ldots \langle E_n, m_n\rangle \}$, $e\in E_i$ and $E_i$ is a match for $m_i$.
Finally, treating these matches as a single sub-graph, assign them to the partition which they share the most incident edges.
This approach would greedily ensure that no edges belonging to motif matching sub-graphs in $G$ ever cross a partition boundary. 
However, it would likely also have the effect of creating highly unbalanced
partition sizes, 
%reducing the potential performance improvements of partitioned
%graphs as queries are unevenly distributed cluster.
%
portentially straining the resources of a single machine, which prompted partitioning in
the first place.

Instead, we rely upon two distinct heuristics for edge assignment, both of which are aware of partition balance.
Firstly, for the case of non-motif-matching edges that are assigned immediately, we use the existing \emph{Linear Deterministic Greedy} (LDG) heuristic \cite{Stanton2012}.
Similar to our naive solution above, LDG seeks to assign edges\footnote{LDG may partition either vertex or edge streams.} to the partition where they have the most incident edges. 
However, LDG also favours partitions with higher residual capacity when assigning edges in order to maintain a balanced number of vertices and edges between each.
Specifically, LDG defines the residual capacity $r$ of a partition $S_i$ in terms of the number of vertices currently in $S_i$, given as $\mathcal{V}(S_i)$, and a partition capacity constraint $C$: $r(S_i) = 1 - \frac{\vert \mathcal{V}(S_i)\vert }{C}$.
When assigning an edge $e$, LDG counts the number of $e$'s incident edges in each partition, given as $N(S_i, e)$, and weights these counts by $S_i$'s residual capacity; $e$ is assigned to the partition with the highest weighted count.
%
%In this way partitions are progressively more penalised as the number of their vertices increases.
%
The full formula for LDG's assignment is: 
\[
	\max_{S_i\in P_k(G)} N(S_i, e)\cdot (1 - \frac{\vert \mathcal{V}(S_i)\vert }{C})
\]
Secondly, for the general case where edges form part of motif matching sub-graphs, we propose a novel heuristic, \emph{equal opportunism}.
Equal opportunism extends ideas present in LDG but, when assigning clusters of motif matching sub-graphs to a single partition as we do in Loom, it has some key advantages.
By construction, given an edge $e$ to be assigned along with its motif matches $M_e = \{\langle E_1, m_1\rangle \ldots \langle E_n, m_n\rangle \}$, the sub-graphs $E_i$ $E_j$ in $M_e$ have significant overlap (e.g. they all contain $e$).
Thus, individually assigning each motif match to potentially different partitions would create many inter-partition edges.
Instead, equal opportunism greedily assigns the match cluster to the single partition with which it shares the most vertices, weighted by each partition's residual capacity.
However, as these vertices and their new motif matching edges may not be traversed with equal likelihood given a workload $Q$, equal opportunism also prioritises the shared vertices which are part of motif matches with higher support in the TPSTry++. 
% Possibly say instead:  "equal oppportunism also prioritises a given partition's overlap with a match cluster if the shared vertices are part of motif matches with partiticularly high support in the TPSTr++."? 

%
Formally, given the motif matches $M_e$ we compute a score for each partition $S_i$ and motif match $\langle E_k, m_k\rangle \in M_e$, which we call a \emph{bid}.
Let $\mathcal{N}(S_i, E_k) = \vert \mathcal{V}(S_i)\cap \mathcal{V}(E_k)\vert $ denote the number of vertices in the edge set $E_k$ (which is itself a graph) that are already assigned to $S_i$\footnote{Note that $\mathcal{N}$ is a generalisation of LDG's function $N$}. 
Additionally, let $supp(m_k)$ refer to the support of motif $m_k$ in the TPSTry++ and recall that $C$ is a capacity constraint defined for each partition.
We define the bid for partition $S_i$ and motif match $\langle E_k, m_k\rangle $ as:
\begin{equation}
bid(S_i, \langle E_k, m_k\rangle ) = \mathcal{N}(S_i, E_k) \cdot (1 - \frac{\vert \mathcal{V}(S_i)\vert }{C}) \cdot supp(m_k)
	\label{eqn:eq_op_bid}
\end{equation}
We could simply assign the cluster of motif matching sub-graphs (i.e. $E_1\cup \ldots \cup E_n$) to the single partition $S_i$ with the highest bid for all motif matches in $M_e$.
However, equal opportunism further improves upon the balance and quality of partitionings produced with this new weighted approach, limiting its greediness using a rationing function we call $l$.
$l(S_i)$ is a number between 0 and 1 for each partition, the size of which is inversely correlated with $S_i$'s size relative to the smallest partition $S_{min} = \min_{S\in P_k(G)} \vert \mathcal{V}(S)\vert $, i.e. if $S_i$ is as small as $S_{min}$ then $l(S_i) = 1$.
Equal opportunism sorts motif matches in $M_e$ in descending order of support, then uses $l(S_i)$ to control both the number of matches used to calculate partition $S_i$'s total bid, and the number of matches assigned to $S_i$ should its total bid be the highest.
This strategy helps create a balanced partitioning by \emph{a)} allowing smaller partitions to compute larger total bids over more motif matches; and \emph{b)} preventing the assignment of large clusters of motif matches to an already large partition.
Formally we calculate $l(S_i)$ as follows: 
\begin{equation}
	l(S_i) = \frac{\vert \mathcal{V}(S_i)\vert }{S_{min}}\cdot \alpha 
\text{~, where~} \alpha = 
\begin{cases}
1, & \vert \mathcal{V}(S_i)\vert  = \vert \mathcal{V}(S_{min})\vert \\
	0, & \vert  \mathcal{V}(S_i)\vert > \vert \mathcal{V}(S_{min})\vert
	\cdot b \\
\alpha, & \text{otherwise}
\end{cases} 
	\label{eqn:eq_op_l}
\end{equation}
where $\alpha$ is a user specified number $0 < \alpha \leq 1$ which controls the aggression
with which $l$ penalises larger partitions and $b$ limits the maximum
imbalance. Throughout this work we use an empirically chosen default of $\alpha
= \frac{2}{3}$ and set the maximum imbalance to $b = 1.1$, emulating
Fennel~\cite{Tsourakakis2012}. 
Given definitions (\ref{eqn:eq_op_bid}) and (\ref{eqn:eq_op_l}), we can now simply state the output of equal oppurtinism for the sorted set of motif matches $M_e$, as:
\begin{equation}
	\max_{S_i\in P_k(G)} \sum_{k=0}^{l(S_i)\cdot \vert M_e\vert} bid(S_i, \langle E_k, m_k\rangle )
	\label{eqn:eq_op_total}
\end{equation}
Note that motif matches in $M_e$ which are not bid on by the winning partition are dropped from the $matchList$ map, as some of their constituent edges (e.g. $e$, which all matches in $M_e$ share) have been assigned to partitions and removed from the sliding window $P_{temp}$.
%
%Remember though, that all sub-graphs of motif matches are also motif matchs; if an $a$-$b$-$c$ motif match is dropped because its $b$-$c$ edge is removed and assigned as part of some other match, there will still exist a separate entry in $matchList$ for the $a$-$b$ edge. 
%
%In other words, given the graph stream window in fig.~\ref{fig:matchList} if the edge $(4,5)$ were to be assigned, the $matchList$ entry $<\{(3,4),(4,5)\}, a\text{-}b\text{-}c >$ would be dropped but the entry $<\{(3,4)\}, a\text{-}b >$ would still exist.
%

% TODO: Find a concise way of saying the below, repeat it a couple of times in various forms
% TODO: Fig
To understand how to the rationing function $l$ improves the quality of equal opportunism's partitioning, not just its balance, consider the following:
Just because an edge $e'$ falls within the motif match set $M_e$ of our assignee $e$, does not necessarily imply that placing them within the same partition is optimal. 
$e'$ could be a member of many other motif matches in $P_{temp}$ besides those in $M_e$, perhaps with higher support in the TPSTry++ (i.e. higher likelihood of being traversed when executing a workload $Q$).
By ordering matches by support and prioritising the assignment of the smaller, higher support motif matches, we often leave $e'$ to be assigned later along with matches to which it is more ``important''.
As an example, consider again the graph and TPSTry++ fragment in Fig.~\ref{fig:matchList}.
If assigning the edge $e_1$ to a partition at the time $t+1$, its \textbf{support ordered} set of motif matches $M_{e_1}$ would be $\langle e_1, m_1\rangle , \langle \{ e_1, e_4 \} , m_3\rangle , \langle \{ e_1, e_5 \} , m_4\rangle $ and $\langle \{ e_1, e_2, e_5 \} , m_6\rangle $.
%
%the sub-graphs $\{(1,2)\}$, $\{(1,2),(2,5)\}$, $\{(1,2),(2,3)\}$ and $\{(1,2),(2,3),(3,4)\}$ for the motifs $a$-$b$, $a$-$b$-$c$, $a$-$b$-$a$, and $a$-$b$-$a$-$b$ respectively.
%
Assume two partitions $S_1$ and $S_2$, where $S_1$ is $33.3\%$ larger than $S_2$ and vertex 2 already belongs to partition $S_1$, whilst all other vertices in the window are as yet unassigned (i.e. this is the first time edges containing them have entered the sliding window).
In this scenario, $S_1$ is guaranteed to win all bids, as $S_2$ contains no vertices from $M_{e_1}$ and therefore $\mathcal{N}(S_2, \_)$ will always equal 0.
However, rather than greedily assign all matches to the already large $S_1$, we
calculate the ration $l$ for $S_1$ as $\frac{1}{1.\overline{33}}\cdot
\frac{1}{1.5} = \frac{1}{2}$, given $\alpha = 1.5$.
In other words, we only assign edges from the first half of $M_{e_1}$ ($\langle e_1, m_1\rangle , \langle \{ e_1, e_4 \} , m_3\rangle $) to $S_1$; edges such as $e_5$ and $e_2$ remain in the window $P_{temp}$. 
Assume an edge $e_6=(4,6)$ subsequently arrives in the graph stream $G$, where vertex 6 already belongs to partition $S_2$ and $e_6$ matches the motif $m_2$ (i.e. has labels $b$-$c$).
If we had already assigned $e_5$ to partition $S_1$ then this would lead to an inter-partition edge which is more likely to be traversed together with $e_5$ than are other edges in $S_1$, given our workload $Q$.
Instead, we compute a match in $P_{temp}$ between $\{e_5, e_6\}$ and the motif $m_3$, and will likely later assign $e_5$ to partition $S_2$.
Within reason, the longer an edge remains in the sliding window, the more of its neighbourhood information we are likely to have access to, the better partitioning decisions we can make for it.

\section{Evaluation} \label{section:evaluation}
Our evaluation aims to demonstrate that Loom achieves high quality partitionings
of several large graphs in a single-pass, streaming manner.
Recall that we measure graph partitioning quality using the number of
inter-partition traversals when executing a realistic workloads of pattern
matching queries over each graph.
Loom consistently produces partitionings of around 20\% superior quality when
compared to those produced by state of the art alternatives:
LDG~\cite{Stanton2012} and Fennel~\cite{Tsourakakis2012}
% 25 ? 
%
Furthermore, Loom partitionings' quality improvement is robust across different numbers of partitions (i.e. a 2-way or a 32-way partitioning).
Finally we show that, like other streaming partitioners, Loom is sensitive to the arrival order of a graph stream, but performs well given a pseudo-adversarial random ordering.
\subsection{Experimental setup}
For each of our experiments, we start by streaming a graph from disk in one of
three predefined orders:
\textbf{Breadth-first:} computed by performing a breadth-first search across all the connected components of a graph;
\textbf{Random:} computed by randomly permuting the existing order of a graph's elements;
and \textbf{Depth-first:} computed by performing a depth-first search across the connected components of a graph.
We choose these stream orderings as they are common to the evaluations of other
graph stream partitioners~\cite{Nishimura2013, Huang2016, Tsourakakis2012,
Stanton2012}, \textbf{including} LDG and Fennel.
Subsequently, we produce 4 separate $k$-way partitionings of this ordered graph stream, using each of the following partitioning approaches for comparison:
\textbf{Hash:} a naive partitioner which assigns vertices and edges to partitions on the basis of a hash function.
As this is the default partitioner used by many existing partition graph databases\footnote{The Titan graph database: \url{http://bit.ly/2ejypXV}}, we use it as a baseline for our comparisons.
\textbf{LDG:} a simple graph stream partitioner with good performance which we extend with our work on Loom.
\textbf{Fennel:} a state-of-the-art graph stream partitioner and our primary point of comparison.
As suggested by Tsourakakis et al, we use the Fennel parameter value $\gamma = 1.5$ throughout our evaluation. 
\textbf{Loom:} our own partitioner which, unless otherwise stated, we invoke
with a window size of 10k edges and a motif support threshold of $40\%$.
%
%Note that although Loom is capable of partitioning arbitary streams of edges, LDG and Fennel expect to receive a stream of vertices along with their adjacency lists.
%%
%In order to facilitate a fair comparison between our systems therefore, all edge
%streams partitioned by Loom are grouped by vertex; i.e. a random order stream might consist of \textit{all} the edges of vertex $v_2$, followed by \textit{all} the edges of vertex $v_7$. 
%
%
% If we want to make more out of our "points of comparison" here is the place to do it.
%

Finally, when each graph is finished being partitioned, we execute the
appropriate query workload over it and count the number of inter-partition
traversals ($ipt$) which occur. 
Note that we avoid implementation dependent measures of partitioning quality
because, as an isolated prototype, Loom is unlikely to exhibit realistic
performance. For instance, lacking a distributed query processing engine, query
workloads are executed over logical partitions during the evaluation. In the
absence of network latency, query response times are meaningless as a measure of
partitioning quality.

All algorithms, data structures, datasets and query workloads are publicly
available\footnote{\label{footnote:loom_repo}The Loom repository: \url{http://bit.ly/2eJxQcp}}.
All our experiments are performed on a commodity machine with a 3.1Ghz Intel i7 CPU and 16GB of RAM.
\subsubsection{Graph datasets}
Remember that the workload-agnostic partitioners which we aim to supersede with
Loom are liable to exhibit poor workload performance when queries focus on
traversing a limited subset of edge types (Sec.~\ref{section:introduction}).
Intuitively, such skewed workloads are more likely over heterogeneous graphs, where there exist a larger number of possible edge types for queries to discern between, e.g. $a$-$a$, $a$-$b$, $a$-$c$\ldots vs just $a$-$a$.  
Thus, we have chosen to test the Loom partitioner over five datasets with a range of different heterogeneities and sizes; three of these datasets are synthetic and two are real-world.  
\begin{figure}
	\centering
	\includegraphics[width=0.9\columnwidth]{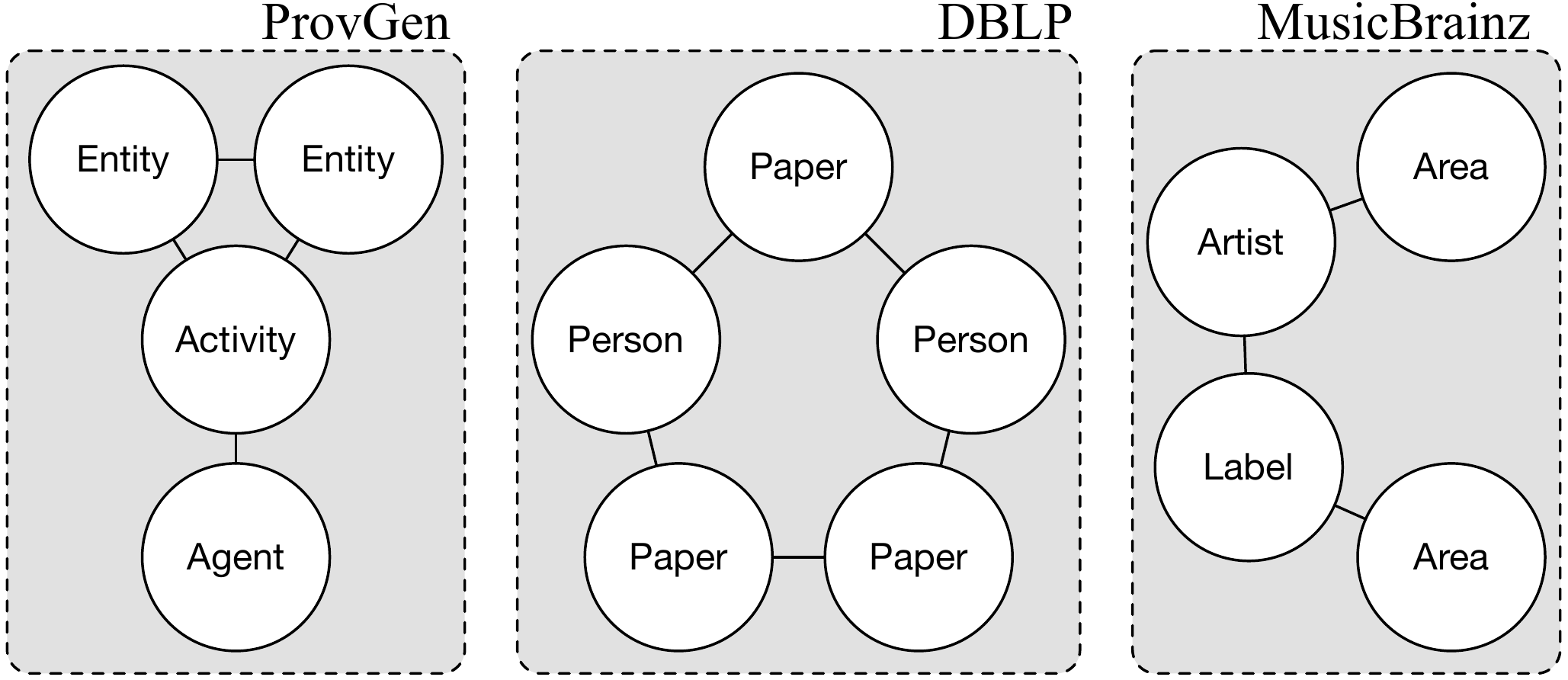}
	\caption{Examples of $q$ for MusicBrainz, DBLP \& ProvGen}\label{fig:eval_query_examples}
\end{figure}
Table~\ref{table:datasets} presents information about each of our chosen datasets, including their size and how heterogeneous they are ($\vert L_V\vert$). 
We use the DBLP, and LUBM datasets, which are well known.
%
%Specifically, we use the LUBM-100, LUBM-4000 datasets.
% TODO: Something about the DBLP version we use here too?
%
MusicBrainz\footnote{The MusicBrainz database: \url{http://bit.ly/1J0wlNR}} is a freely available database of curated music metadata, with vertex labels such as \emph{Artist}, \emph{Country}, \emph{Album} and \emph{Label}.
ProvGen\cite{Firth2014} is a synthetic generator for PROV metadata~\cite{w3c-prov-dm}, which records detailed provenance for digital artifacts.

%
%A PROV graph naturally has 3 vertex labels: \emph{Entity} (data), \emph{Activity} (execution of a process which acts upon data) and \emph{Agent} (people or systems responsible for activities).
%
\subsubsection{Query workloads}
For each dataset we must propose a representative query workload to execute so that we may measure partitioning quality in terms of $ipt$.
Remember that a query workload consists of a set of distinct query patterns along with a frequency for each (Sec.~\ref{section:preliminaries}). 
The LUBM dataset provides a set of query patterns which we make use of.
For every other dataset, however, we define a small set of common-sense queries
which focus on discovering implicit relationships in the graph, such as
potential collaboration between authors or artists~\footnote{If possible,
workloads are drawn from the literature, e.g. common PROV
queries~\cite{dey2013implementing}}. 
The full details of these query patterns are elided for
space\footnoteref{footnote:loom_repo}, however Fig.~\ref{fig:eval_query_examples} presents some examples.
\begin{table}
	\small
	\centering
	\setlength\tabcolsep{2pt} % default value: 6pt
	\begin{tabular}{| l | l | l | l | l | l |}
		\hline
		Dataset & $\sim \vert V\vert$ & $\sim \vert E\vert$ & $\vert L_V\vert$ & Real & Description \\ \hline
		DBLP  & 1.2M & 2.5M & 8 & Y & Publications \& citations \\ \hline
		ProvGen & 0.5M & 0.9M & 3 & N & Wiki page provenance \\ \hline
		MusicBrainz & 31M & 100M & 12 & Y & Music records metadata \\ \hline
		LUBM-100 & 2.6M & 11M & 15 & N & University records \\ \hline
		LUBM-4000 & 131M & 534M & 15 & N & University records \\ \hline
		%DBPedia & 5.5M & 20M & 2 & Y & Wikipedia categories \\
	\end{tabular}
	\caption{Graph datasets, incl.  size \& heterogeneity}\label{table:datasets}
	\vspace{-2em}
\end{table}
%
% Where possible, we design queries based upon uses cases for each dataset in
% the literature, for example .... prov query reference from SigMod.
%
%Note that our current prototype of the Loom partitioner assumes that the
%frequencies of queries in a given workload are fixed, and known prior to
%starting the partitioning of a graph.
%
Note that whilst the TPSTry++ may be trivially updated to account for change in
the frequencies of workload queries (Sec.~\ref{section:motif_mine}), our
evaluation of Loom assumes that said frequencies are fixed and known \textit{a
priori}.
Recall that, for online databases, we argue this is a realistic
assumption (Sec.~\ref{section:introduction}).
However, more complete tests with changing workloads are an important area for
future work.
%
%However, unless stated otherwise, we run each experiment multiple times,
%randomizing the workload query frequencies each time.
%
%We then present the averages $\%$ reduction in $ipt$ achieved, relative to our
%baseline Hash partitioning, by each partitioner and for each dataset.
%
\subsection{Comparison of systems} 
\begin{figure*}
	\centering
	\subfloat[Random order]{
		\includegraphics[width=0.55\columnwidth]{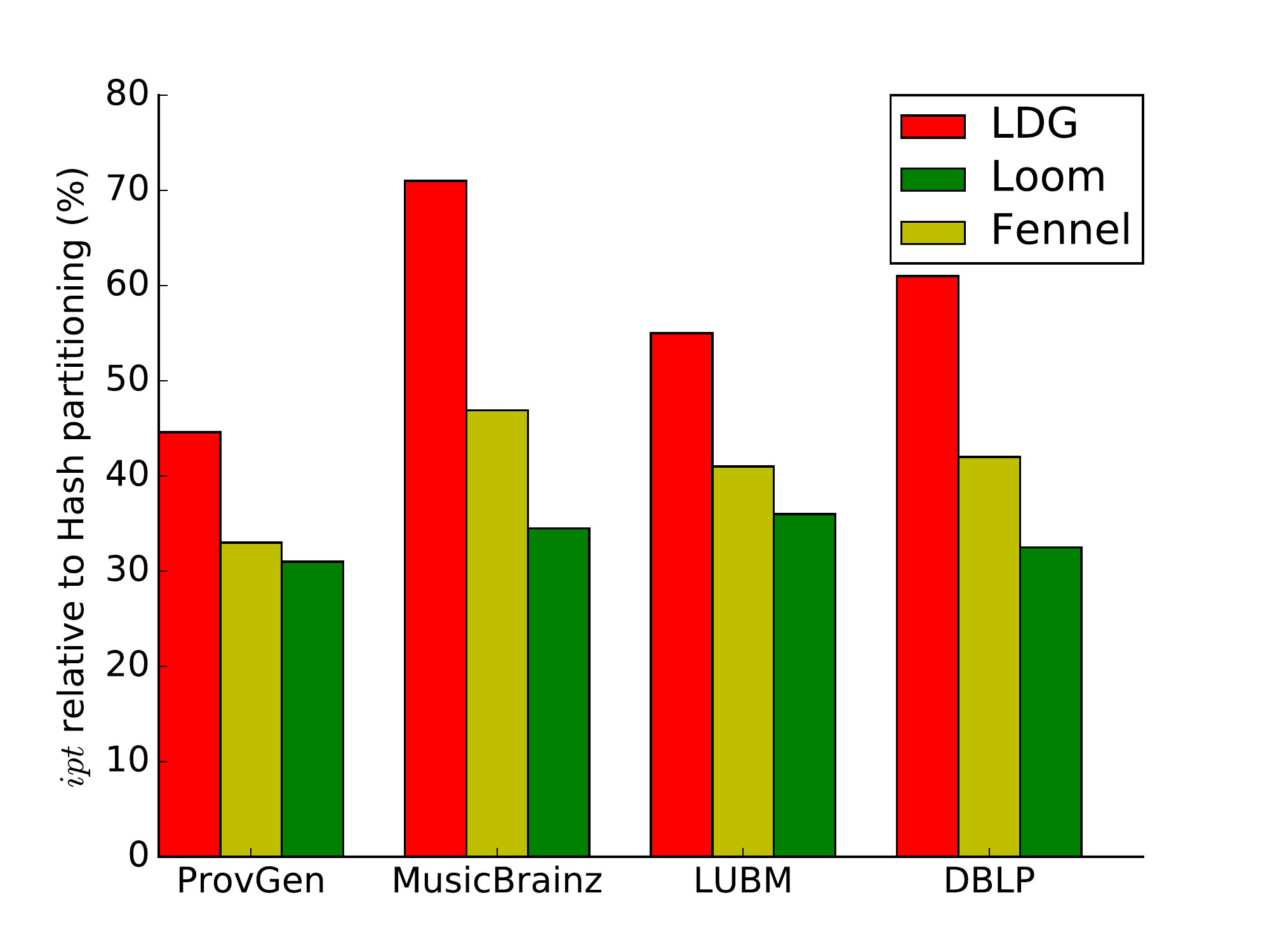}
		\label{fig:rel-rand-k8}%
	}
	\qquad
	\subfloat[Breadth-first order]{
		\includegraphics[width=0.55\columnwidth]{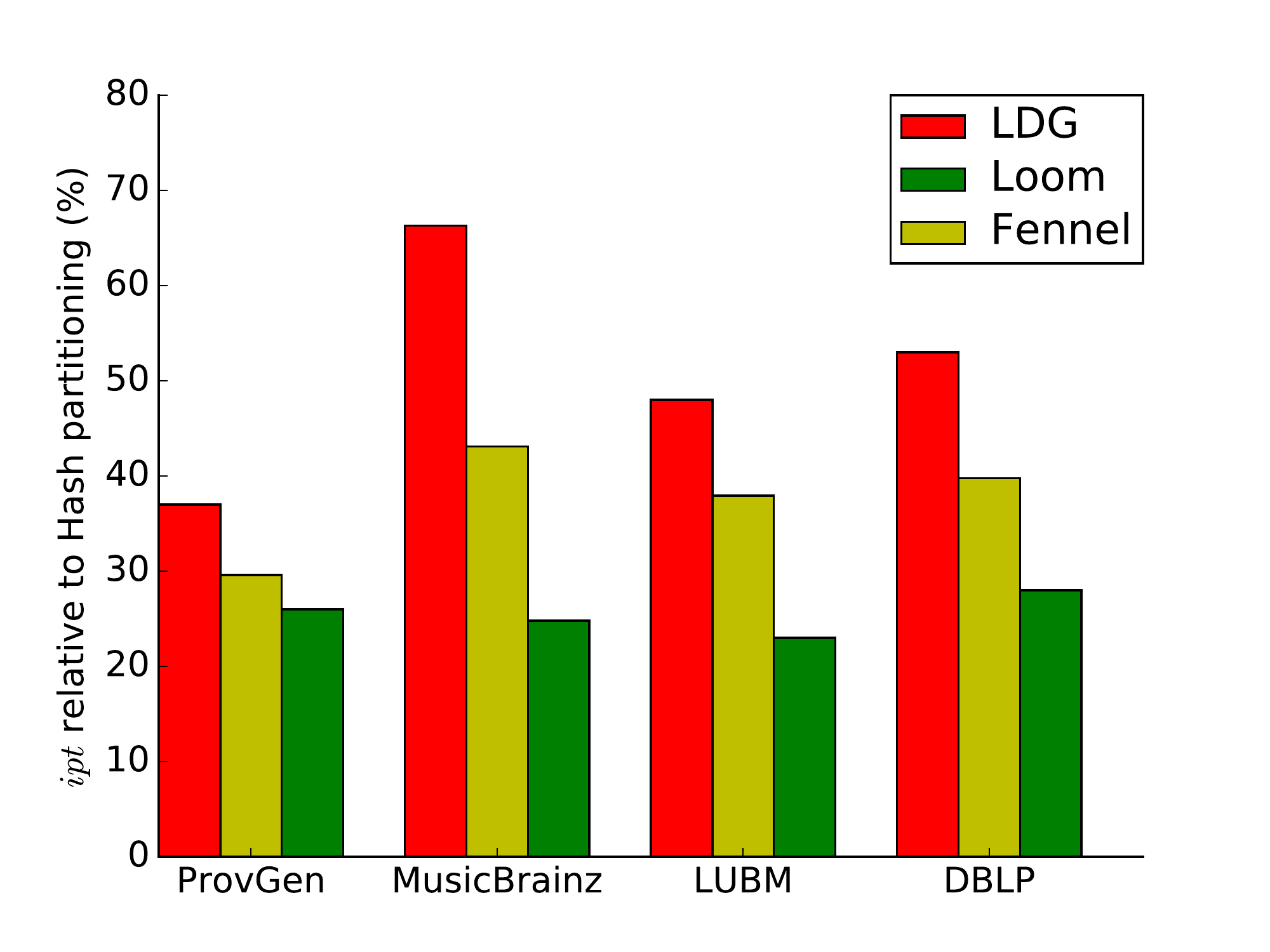}
		\label{fig:rel-bfs-k8-b}%
	}
	\qquad
	\subfloat[Depth-first order]{
		\includegraphics[width=0.55\columnwidth]{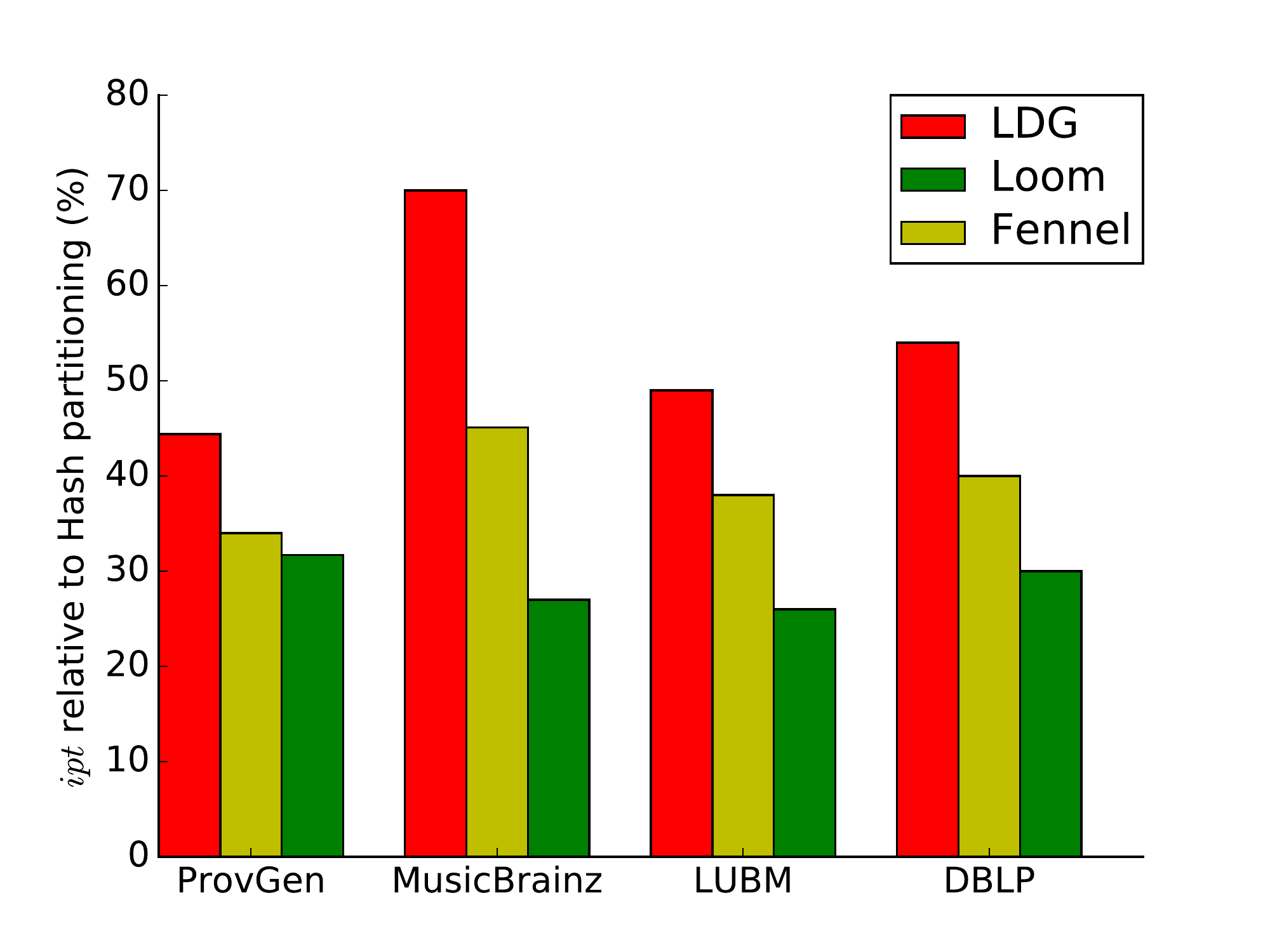}
		\label{fig:rel-dfs-k8}%
	}
%	\caption{$ipt$, relative to our Hash partitioning baseline, when executing $Q$ over \textbf{8-way} partitionings of graph streams in multiple orders.}
	\caption{$ipt$ \%, vs. Hash,  when executing $Q$ over \textbf{8-way} partitionings of graph streams in multiple orders.}
	\label{fig:rel-all-k8}
\end{figure*}
\begin{figure*}
	\centering
	\subfloat[$k = 2$]{
		\includegraphics[width=0.55\columnwidth]{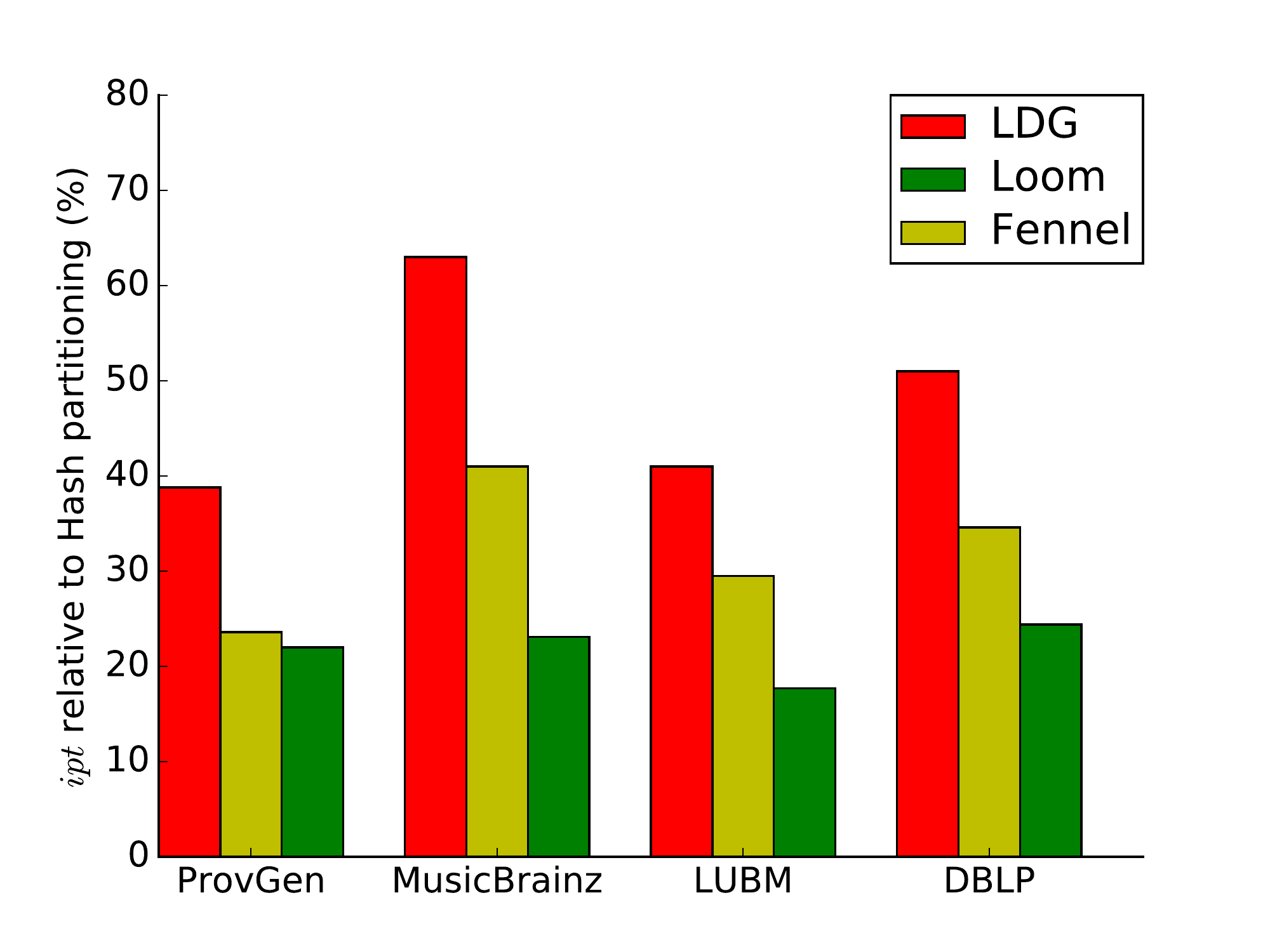}
		\label{fig:rel-bfs-k2}%
	}
	\qquad
	\subfloat[$k = 8$]{
		\includegraphics[width=0.55\columnwidth]{rel-bfs-k8}
		\label{fig:rel-bfs-k8}%
	}
	\qquad
	\subfloat[$k = 32$]{
		\includegraphics[width=0.55\columnwidth]{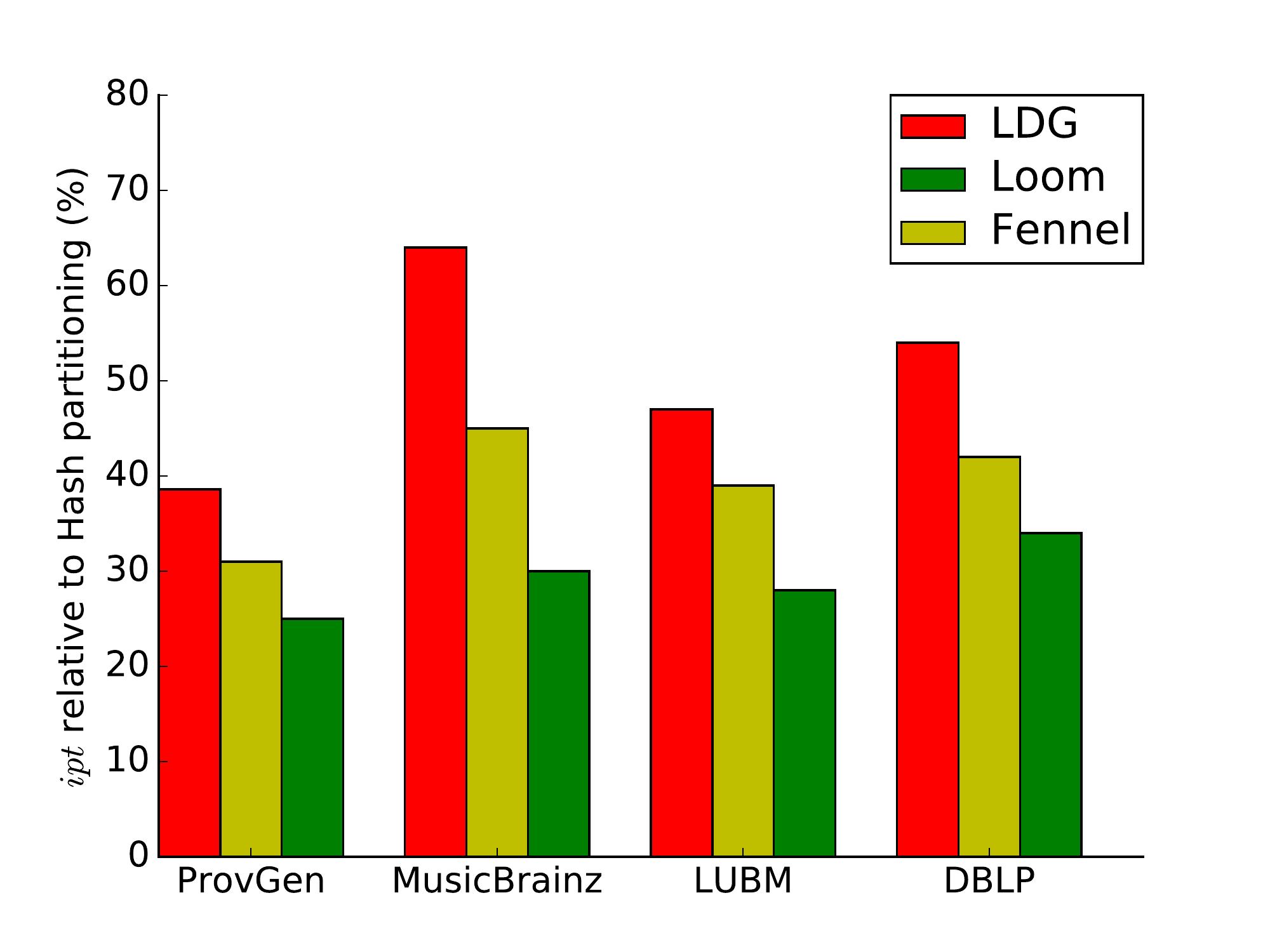}
		\label{fig:rel-bfs-k32}%
	}
	%\caption{$ipt$, relative to our Hash partitioning baseline, when executing $Q$ over multiple $k$-way partitionings of \textbf{breadth-first} graph streams .}
	\caption{$ipt$ \%, vs. Hash, when executing $Q$ over multiple $k$-way
	partitionings of \textbf{breadth-first} graph streams.}
	\label{fig:rel-bfs-all-k}
	% Make all text in graphs bigger.
\end{figure*}
Figures \ref{fig:rel-all-k8} and~\ref{fig:rel-bfs-all-k} present the improvement
in partitioning quality achieved by Loom and each of the comparable systems we
desribe above.
Initially, consider the experiment depicted in Fig.~\ref{fig:rel-all-k8}.
We partition ordered streams of each of our first 4 graph
datasets\footnote{Excluding LUBM-4000} into 8-way partitionings, using the
 approaches described above, then execute each dataset's query
workload over the appropriate partitioning.
The absolute number of inter-partition traversals ($ipt$) suffered when querying
each dataset varies significantly.
Thus, rather than represent these results directly, in Fig.~\ref{fig:rel-all-k8}
(and~\ref{fig:rel-bfs-all-k}) we present the results for each approach as
\textbf{relative} to the results for Hash; i.e. how many $ipt$ did
a partitioning suffer, \textbf{as a percentage} of those suffered by the Hash
partitioning of the same dataset.
As expected, the naive hash partitioner performs poorly: it produces partitionings which suffer twice as many inter-partition traversals, on average, when compared to partitionings produced by the next best system (LDG). % Bother with this sentence?
%
% Talk about the performance of LDG, compare it to Loom.
Whilst the LDG partitioner does achieve around a $55\%$ reduction in $ipt$ vs our Hash baseline, its produces partitionings of consistently poorer quality than those of Fennel and Loom.
Although both LDG and Fennel optimise their partitionings for the balanced min. edge-cut goal (Sec.~\ref{section:introduction}), Fennel is the more effective heuristic, cutting around $25\%$ fewer edges than LDG for small numbers of partitions (including $k=8$)~\cite{Tsourakakis2012}.
Intuitively, the likelihood of \textit{any} edge being cut is a coarse proxy for the likelihood of a query $q\in Q$ traversing a cut edge.
This explains the disparity in $ipt$ scores between the two systems. 
%

% Talk about performance of Fennel, compare it to Loom.
Of more interest is comparing the quality of partitionings produced by Fennel and Loom. 
Fig.~\ref{fig:rel-all-k8} clearly demonstrates that Loom offers a significant improvement in partitioning quality over Fennel, given a workload $Q$. 
% Say more here, perhaps about underlining the importance of workload sensitivity.
% Do we want to say this or do we want a slight loss to Fennel on Random DBPedia? Thinking slight loss is more believable and gives us something to talk about, but want to work out what the categories are first. 
Loom's reduction in $ipt$ relative to Fennel's is present across all datasets and stream orders, however it is particularly pronounced over ordered streams of more heterogeneous graphs; e.g. MusicBrainz in subfigure~\ref{fig:rel-bfs-k8}, where Loom's partitioning suffers from $42\%$ fewer $ipt$ than Fennel's. 
This makes sense because, as mentioned, pattern matching workloads are more likely to exhibit skew over heterogeneous graphs, where query graphs $G_q$ contain a, potentially small, subset of the possible vertex labels.
%
%Additionally, when partitioning breadth and depth-first ordered graph streams, if the average degree of the graph is sufficiently low then  all instances of a motif match are likely to be detected within even a moderately sized Loom window.
%
%For example, if the degree of a given vertex $a$ is 10, and the largest motif contains 5 edges, then a breadth-first traversal of a 100k edges\footnote{i.e. a window of 100k edges over a breadth-first stream.} from $a$ is guaranteed to include all the motif matches which contain it.
%
%This allows Loom's equal opportunism heuristic to make the best possible allocation decisions for groups of vertices.
%
Across all the experiments presented in Fig.~\ref{fig:rel-all-k8}, the median range of Loom's $ipt$ reduction relative to Fennel's is $20-25\%$. 
% TODO: Double check this.
% TODO: Talk here about the scalability with the number of partitions...
Additionally, Fig.~\ref{fig:rel-bfs-all-k} demonstrates that this improvement is consistent for different numbers of partitions.
%
%\authornote{HF}{The below is really confusing, so lets rewrite a little bit.}
%
As the number of partitions $k$ grows, there is a higher probability that vertices belonging to a motif match are assigned across multiple partitions.
% in order to maintain balance?
This results in an increase of \textit{absolute} $ipt$ when executing $Q$ over a Loom partitioning.
%
%\authornote{HF}{Still need to fix the sentence below, rewrite the point about average degree to talk about degree distribution, make sure that the window fig supports that and that the bar charts support the windows.}
However, increasing $k$ actually increases the probability that any two vertices which share an edge are split between partitions, thus reducing the quality of Hash, LDG and Fennel partitionings as well. 
As a result, the difference in \textit{relative} $ipt$ is largely consistent between all 4 systems.
% Technically not true - Fennel gets closer to LDG, whilst we don't. Can we make that point?
%

%
On the other hand, neither Fig.~\ref{fig:rel-all-k8}, nor Fig.~\ref{fig:rel-bfs-all-k}, present the runtime costs of producing a partitioning. 
%
%Table~\ref{table:perf} presents how long (in seconds) each partitioner takes to
%partition our three largest graph datasets. 
Table~\ref{table:perf} presents how long (in ms) each partitioner takes to
partition 10k edges. 
%
%We find that Loom is around a factor of 3 slower than either LDG or Fennel, 
Whilst all 3 algorithms are capable of partitioning many 10s of thousands of
edges per second, we do find that Loom is slower than LDG and Fennel by an
average factor of 2-3.
%whose performance are similar.
%
This is likely due to the more complex map-lookup and pattern-matching logic
performed by Loom, or a nascent implementation.
%upon the arrival of each element of the graph stream. 
%
The runtime performance of Loom varies depending on the query workload $Q$ used
to generate the TPSTry++ (Sec.~\ref{section:motif_mine}), therefore the
performance figures presented in Table~\ref{table:perf} are averaged across many
different $Q$.
The minimum slowdown factor observed between Loom and Fennel was 1.5, the maximum 7.1.
Note that popular non-streaming partitioner Metis~\cite{Karypis1998a} is around
13 times slower than Fennel for large graphs~\cite{Tsourakakis2012}.
\begin{table}
	\small
	\centering
	\setlength\tabcolsep{2pt} % default value: 6pt
	\begin{tabular}{| l | l | l | l | l |}
		\hline
		Dataset & LDG (ms) & Fennel (ms) & Loom (ms) & Hash (ms) \\ \hline
		DBLP & 91 & 96 & 235 & 28 \\ \hline
		ProvGen & 144 & 146 & 240 & 33 \\ \hline
		MusicBrainz & 48 & 52 & 129 & 18 \\ \hline
		LUBM-100 & 47 & 51 & 147 & 22 \\ \hline
		LUBM-4000 & 45 & 49 & 138 & 16 \\ \hline
	\end{tabular}
	\caption{Time to partition 10k edges}\label{table:perf}
	\vspace{-2em}
\end{table}

We contend that this performance difference is unlikely to be an issue in an
online setting for two reasons. 
Firstly, most production databases do not support more than around 10k
transactions per second (TPS)~\cite{Lee2008}.
Secondly, it is considered exceptional for even applications such as twitter to
experience $>$30k-40k TPS~\footnote{Tweets per second in 2013:
\url{http://bit.ly/2hQH5JJ}}. 
Meanwhile, the lowest partitioning rate exhibited by Loom in
Table~\ref{table:perf} is equivalent to \textasciitilde~42k edges per second, the highest 72k. 
%We contend that such speed differences are not relevant for two reasons.
%%
%Firstly, in offline, graph-analytical systems the execution time of analytics workloads exceeds even complex partitioning approaches by orders of magnitude\cite{Xu2014}.
%%
%%\authornote{HF}{Again, I know we're short on space but I think we need to really hit this home with following through?}
%Therefore, if increasing partitioning time 3-fold decreases the workload execution time by $25\%$ then the overall system runtime is still reduced.
%%
%Secondly, in online systems such speed differences are only noticable for a graph arrival rate of multiple hundreds of thousands of graph elements per second, which even systems such as twitter do not experience~\footnote{Tweets per second in 2013: \url{http://bit.ly/2hQH5JJ}}.
%

%
Note that Figures~\ref{fig:rel-all-k8} and \ref{fig:rel-bfs-all-k} do not present the relative $ipt$ figures for the LUBM-4000 dataset.
This is because measuring relative $ipt$ involves reading a partitioned graph into memory, which is beyond the constraints of our present experimental setup.
However, we include the LUBM-4000 dataset in Table~\ref{table:perf} to demonstrate that, as a streaming system, Loom is capable of partitioning large scale graphs.
Also note that none of the figures present partitioning imbalance as this is
broadly similar between all approaches and datasets~\footnote{Except Hash, which
is balanced.}, with LDG varying between $1\% - 3\%$, Loom and Fennel between
$7\%$ and their maximum imbalance of $10\%$ (Sec.~\ref{section:motif_alloc}).
% (?) Plan to talk about the performance of Metis, compare it to Loom. 

% Talk about the graphs we have and how we compare to Fennel and LDG
% If you get the change, talk about how we compare (and Fennel and LDG if you like) to METIS, possibly over just the one dataset. This is unlikely to be possible
% If possible, bring up a comparison to the number of edges cut (which we would have to instrument, also another new graph?) to make the point that we are successfully prioritisiing sub-graphs based on their ``relevance'' to a given query workload, but also make the point that (because we are a streaming partitioner and do not consider vertices for reassignment) this makes us sensitive to workload change
% This should be the largest section, because we should also have subsubsections for "Scalability with number of partitions $k$" and "Sensitivity to stream order".
% 
%\subsubsection{Sensitivity to stream order}
%

%

%
%\subsubsection{Scalability with number of partitions $k$}
%
%
\subsection{Effect of stream order and window size}
\begin{figure}
	\centering
	\includegraphics[width=0.9\columnwidth]{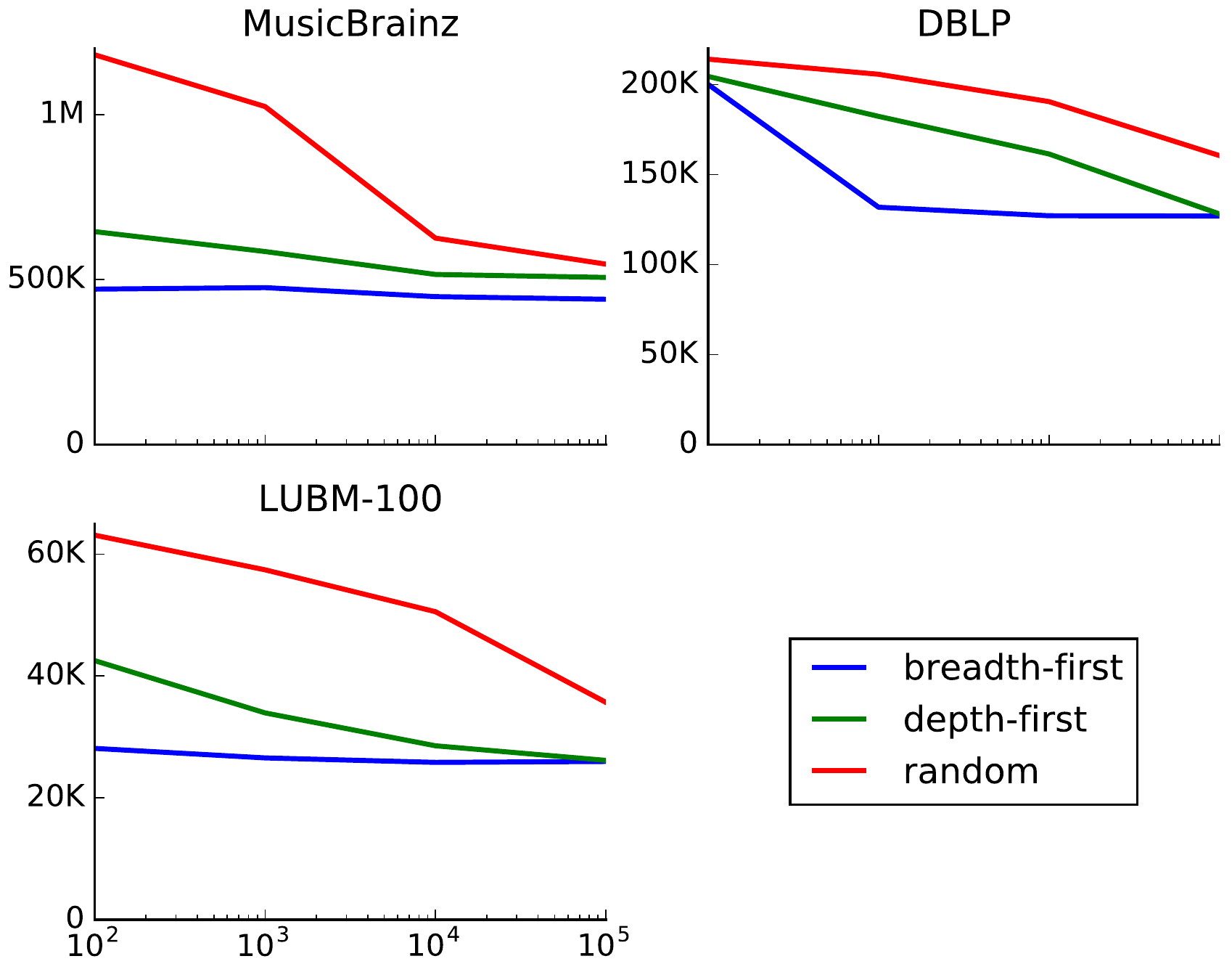}
	\caption{$ipt$ (\textit{y-axis}) when executing $Q$ over Loom partitionings with multiple window sizes $t$(\textit{x-axis})}
	\label{fig:ipt-vs-t}
	\vspace{-1em}
\end{figure}

Fig.~\ref{fig:rel-all-k8} indicates that Loom is sensitive to the ordering of its given graph stream.
In fact, subfigure \ref{fig:rel-all-k8}a shows Loom achieve a smaller reduction in $ipt$ over Fennel and LDG, than in \ref{fig:rel-all-k8}b and  \ref{fig:rel-all-k8}c. 
Specifically Loom achieves a $42\%$ greater reduction in relative $ipt$ than Fennel given a breadth-first stream of the MusicBrainz graph, but only a $26\%$ when the stream is ordered randomly, despite Fennel and LDG also being sensitive to stream ordering~\cite{Stanton2012, Tsourakakis2012}.
This implies that Loom is particularly sensitive to random orderings: edges which are close to one another in the graph may not be close in the graph stream, resulting in Loom detecting fewer motif matching subgraphs in its stream window. 
Intuitively, this sensitivity can be ameliorated by increasing the size of Loom's window, as shown in Fig.~\ref{fig:ipt-vs-t} 
%
%For example, if the degree of a given vertex $a$ is 8, and the largest motif contains 5 edges, then a breadth-first traversal of a $8^5$ edges (i.e. window size $t\approx 33k$) from $a$ is likely to include all the motif matches which contain it.
%
As Loom's window grows, so does the probability that clusters of motif matching subgraphs will occur within it.
This allows Loom's equal opportunism heuristic to make the best possible allocation decisions for the subgraph's constituent vertices.
Indeed, the number of $ipt$ suffered by Loom partitionings improves significantly, by as much as $47\%$, as the window size grows from 100 to 10k.
However, increasing the window size past $10$k clearly has little effect on $ipt$ suffered to execute $Q$ if your graph stream is ordered.
The exact impact of increasing Loom's window size depends upon the degree distribution of the graph being partitioned.
However, to gain an intuition consider the naive case of a graph with a uniform average vertex degree of 8, along with a TPSTry++ whose largest motif contains 4 edges. 
In this case, a breadth-first traversal of $8^4$ edges from a vertex $a$ (i.e. window size $t\approx 4k$) is highly likely to include all the motif matches which contain $a$.
%To understand this, consider that if the degree of a given vertex $a$ is 8, and the TPSTry++'s largest motif contains 4 edges, a breadth-first traversal of a $8^5$ edges (i.e. window size $t\approx 33k$) from $a$ is highly likely to include all the motif matches which contain it.
%
%The highest average degrees amongst our datasets are MusicBrainz ($6.4$) and LUBM ($8.2$).
%
%
Regardless, Fig.~\ref{fig:ipt-vs-t} might seem to suggest that Loom should run with the largest window size possible.
However, besides the additional computational cost of detecting more motif matches, remember that Loom's window constitutes a temporary partition (sec.~\ref{section:motif_match}).
If there exist many edges between other partitions and $P_{temp}$, then this may itself be a source of $ipt$ and poor query performance. 
%
% Regardlesis
%Here we talk about the effect of changing window size on the different ordered graph partitionings
%Also talk about the effect of window size on wall clock time if we get the chance.
%
%
% Probably not a huge amount to say here, other than a small table and some comments. 
%
%
%

\section{Conclusion} \label{section:conclusion}
In this paper, we have presented Loom: a practical system for producing $k$-way
partitionings of online, dynamic graphs, which are optimised for a given workload of pattern matching queries $Q$.
% of a given query workload.
%
%By constructing a generalised trie datastructure from the query graphs in $Q$, we are able to identify sub-graphs common to many of those queries: \emph{motifs}.
%
%Using an efficient form of pattern matching over the stream of edge additions which constitute a growing graph, we are able to detect matches for these motifs as they form.
%
%Finally, using a novel heuristic, we assign the majority of motif matches within single partitions, whilst maintaining partition balance.
%
Our experiments indicate that 
%, with this strategy, 
Loom significantly reduces the number of inter-partition traversals ($ipt$) required when executing $Q$ over its partitionings, relative to state of the art (workload agnostic) streaming partitioners.
%
% In a distributed environment, lower $ipt$ corresponds to improved performance when executing a workload.
% TODO: review removing the below two paragraphs and bringing in some future work.
%Graph partitioning's produced using Loom consistently exhibit over $70\%$ fewer $ipt$, relative to our baseline of a hash partitioning, and around $25\%$ fewer relative to Fennel, a state-of-the-art streaming graph partitioner.
%
%Furthermore, Loom's partitioning's are both well balanced in the number of vertices (all partitions within $10\%$), and efficient to produce, within a factor 4 of the Fennel and LDG heuristics which are known to be very fast \cite{Tsourakakis2012}. %\cite{Huang2016} instead?
%
%

There are several ways in which we intend to expand our current work on Loom.
In particular, as a workload sensitive technique, Loom generates partitionings which are vulnerable to workload change over time.  
In order to address this we must integrate Loom with an existing, workload
sensitive, graph \textbf{re}-partitioner~\cite{Firth2017} or consider some form of restreaming approach~\cite{Huang2016}.
%
%For instance, we do not consider any sort of replication for the vertices and edges of the graph being partitioned, whilst LEOPARD~\cite{Huang2016} and Peng et al.~\cite{Peng2016} use replication to great effect. 
%%
%Using limited replication, we might be able to place more motif matches within single partitions, rather than splitting them with our equal opportunism heuristic (Sec.~\ref{section:motif_alloc}). 
%%
%Additionally, due to our approaches reliance upon graph pattern matching in a single stream window, Loom is single threaded. 
%%
%The ability to have multiple parallel instances of Loom assigning motif matches to the same graph partitioning would doubtless increase the systems applicaiblity to large scale, online deployment.
%

% ensure same length columns on last page (might need two sub-sequent latex runs)
%\balance
\citestyle{acmnumeric}
\bibliographystyle{ACM-Reference-Format}
\small
\bibliography{loom-edbt}


\begin{thebibliography}{00}

%%% the Web address.

\ifx \showCODEN    \undefined \def \showCODEN     #1{\unskip}     \fi
\ifx \showDOI      \undefined \def \showDOI       #1{#1}\fi
\ifx \showISBNx    \undefined \def \showISBNx     #1{\unskip}     \fi
\ifx \showISBNxiii \undefined \def \showISBNxiii  #1{\unskip}     \fi
\ifx \showISSN     \undefined \def \showISSN      #1{\unskip}     \fi
\ifx \showLCCN     \undefined \def \showLCCN      #1{\unskip}     \fi
\ifx \shownote     \undefined \def \shownote      #1{#1}          \fi
\ifx \showarticletitle \undefined \def \showarticletitle #1{#1}   \fi
\ifx \showURL      \undefined \def \showURL       {\relax}        \fi
% invisible to TeX
\providecommand\bibfield[2]{#2}
\providecommand\bibinfo[2]{#2}
\providecommand\natexlab[1]{#1}
\providecommand\showeprint[2][]{arXiv:#2}

\bibitem[\protect\citeauthoryear{Andreev and Racke}{Andreev and Racke}{2006}]%
        {Andreev2006}
\bibfield{author}{\bibinfo{person}{K. Andreev} {and} \bibinfo{person}{H.
  Racke}.} \bibinfo{year}{2006}\natexlab{}.
\newblock \showarticletitle{{Balanced Graph Partitioning}}.
\newblock \bibinfo{journal}{{\em Theory of Computing Systems\/}}
  \bibinfo{volume}{39}, \bibinfo{number}{6} (\bibinfo{year}{2006}),
  \bibinfo{pages}{929--939}.
\newblock
\showISSN{1432-4350}


\bibitem[\protect\citeauthoryear{Chevalier and Pellegrini}{Chevalier and
  Pellegrini}{2008}]%
        {Chevalier2008a}
\bibfield{author}{\bibinfo{person}{C. Chevalier} {and} \bibinfo{person}{F.
  Pellegrini}.} \bibinfo{year}{2008}\natexlab{}.
\newblock \showarticletitle{{PT-Scotch: A tool for efficient parallel graph
  ordering}}.
\newblock \bibinfo{journal}{{\it Parallel Comput.}} \bibinfo{volume}{34},
  \bibinfo{number}{6-8} (\bibinfo{year}{2008}), \bibinfo{pages}{318--331}.
\newblock
\showISSN{01678191}


\bibitem[\protect\citeauthoryear{Choudhury, Holder, Chin,
  et~al\mbox{.}}{Choudhury et~al\mbox{.}}{2015}]%
        {Choudhury2015a}
\bibfield{author}{\bibinfo{person}{S. Choudhury}, \bibinfo{person}{L. Holder},
  \bibinfo{person}{G. Chin}, {et~al\mbox{.}}} \bibinfo{year}{2015}\natexlab{}.
\newblock \showarticletitle{{A Selectivity based approach to Continuous Pattern
  Detection in Streaming Graphs}}.
\newblock \bibinfo{journal}{{\em Proc. EDBT\/}} (\bibinfo{year}{2015}),
  \bibinfo{pages}{157--168}.
\newblock
\showISBNx{9783893180677}


\bibitem[\protect\citeauthoryear{Curino, Jones, Zhang, et~al\mbox{.}}{Curino
  et~al\mbox{.}}{2010}]%
        {Curino2010}
\bibfield{author}{\bibinfo{person}{C. Curino}, \bibinfo{person}{E. Jones},
  \bibinfo{person}{Y. Zhang}, {et~al\mbox{.}}} \bibinfo{year}{2010}\natexlab{}.
\newblock \showarticletitle{{Schism}}.
\newblock \bibinfo{journal}{{\em Proc. VLDB\/}} \bibinfo{volume}{3},
  \bibinfo{number}{1-2} (\bibinfo{year}{2010}), \bibinfo{pages}{48--57}.
\newblock
\showISSN{21508097}


\bibitem[\protect\citeauthoryear{Dey, Cuevas-Vicentt{\'{i}}n, K{\"{o}}hler,
  et~al\mbox{.}}{Dey et~al\mbox{.}}{2013}]%
        {dey2013implementing}
\bibfield{author}{\bibinfo{person}{S. Dey}, \bibinfo{person}{V.
  Cuevas-Vicentt{\'{i}}n}, \bibinfo{person}{S. K{\"{o}}hler}, {et~al\mbox{.}}}
  \bibinfo{year}{2013}\natexlab{}.
\newblock \showarticletitle{{On implementing provenance-aware regular path
  queries with relational query engines}}. In \bibinfo{booktitle}{{\em Proc.
  EDBT/ICDT Workshops}}. \bibinfo{pages}{214--223}.
\newblock


\bibitem[\protect\citeauthoryear{Firth and Missier}{Firth and Missier}{2014}]%
        {Firth2014}
\bibfield{author}{\bibinfo{person}{H. Firth} {and} \bibinfo{person}{P.
  Missier}.} \bibinfo{year}{2014}\natexlab{}.
\newblock \showarticletitle{{ProvGen: Generating Synthetic PROV Graphs with
  Predictable Structure}}. In \bibinfo{booktitle}{{\em Proc. IPAW}}.
  \bibinfo{pages}{16--27}.
\newblock
\showISBNx{978-3-319-16461-8}


\bibitem[\protect\citeauthoryear{Firth and Missier}{Firth and Missier}{2016}]%
        {Firth2016c}
\bibfield{author}{\bibinfo{person}{Hugo Firth} {and} \bibinfo{person}{Paolo
  Missier}.} \bibinfo{year}{2016}\natexlab{}.
\newblock \showarticletitle{Workload-aware Streaming Graph Partitioning.}. In
  \bibinfo{booktitle}{{\em Proc. EDBT/ICDT Workshops}}.
\newblock


\bibitem[\protect\citeauthoryear{Firth and Missier}{Firth and Missier}{2017}]%
        {Firth2017}
\bibfield{author}{\bibinfo{person}{H. Firth} {and} \bibinfo{person}{P.
  Missier}.} \bibinfo{year}{2017}\natexlab{}.
\newblock \showarticletitle{{TAPER: query-aware, partition-enhancement for
  large, heterogenous graphs}}.
\newblock \bibinfo{journal}{{\em Distributed and Parallel Databases\/}}
  \bibinfo{volume}{35}, \bibinfo{number}{2} (\bibinfo{year}{2017}),
  \bibinfo{pages}{85--115}.
\newblock
\showISSN{15737578}


\bibitem[\protect\citeauthoryear{Gupta, Satuluri, Grewal, et~al\mbox{.}}{Gupta
  et~al\mbox{.}}{2014}]%
        {Gupta2014}
\bibfield{author}{\bibinfo{person}{P. Gupta}, \bibinfo{person}{V. Satuluri},
  \bibinfo{person}{A. Grewal}, {et~al\mbox{.}}}
  \bibinfo{year}{2014}\natexlab{}.
\newblock \showarticletitle{{Real-time twitter recommendation}}.
\newblock \bibinfo{journal}{{\em Proc. VLDB\/}} \bibinfo{volume}{7},
  \bibinfo{number}{13} (\bibinfo{year}{2014}), \bibinfo{pages}{1379--1380}.
\newblock
\showISSN{21508097}


\bibitem[\protect\citeauthoryear{Huang and Abadi}{Huang and Abadi}{2016}]%
        {Huang2016}
\bibfield{author}{\bibinfo{person}{J. Huang} {and} \bibinfo{person}{D. Abadi}.}
  \bibinfo{year}{2016}\natexlab{}.
\newblock \showarticletitle{{LEOPARD : Lightweight Edge-Oriented Partitioning
  and Replication for Dynamic Graphs}}.
\newblock \bibinfo{journal}{{\em Proc. VLDB\/}} \bibinfo{volume}{9},
  \bibinfo{number}{7} (\bibinfo{year}{2016}), \bibinfo{pages}{540--551}.
\newblock
\showISSN{2150-8097}


\bibitem[\protect\citeauthoryear{Jiang, Coenen, and Zito}{Jiang
  et~al\mbox{.}}{2004}]%
        {Jiang2004a}
\bibfield{author}{\bibinfo{person}{C. Jiang}, \bibinfo{person}{F. Coenen},
  {and} \bibinfo{person}{M. Zito}.} \bibinfo{year}{2004}\natexlab{}.
\newblock \showarticletitle{{A Survey of Frequent Subgraph Mining Algorithms}}.
\newblock \bibinfo{journal}{{\em The Knowledge Engineering Review\/}}
  \bibinfo{volume}{000} (\bibinfo{year}{2004}), \bibinfo{pages}{1--31}.
\newblock
\showISBNx{0000000000000}
\showISSN{0269-8889}


\bibitem[\protect\citeauthoryear{Jindal and Dittrich}{Jindal and
  Dittrich}{2012}]%
        {Jindal2012}
\bibfield{author}{\bibinfo{person}{A. Jindal} {and} \bibinfo{person}{J.
  Dittrich}.} \bibinfo{year}{2012}\natexlab{}.
\newblock \showarticletitle{{Relax and let the database do the partitioning
  online}}.
\newblock In \bibinfo{booktitle}{{\em Enabling Real-Time Business
  Intelligence}}. \bibinfo{pages}{65--80}.
\newblock
\showISBNx{9783642334993}
\showISSN{18651348}


\bibitem[\protect\citeauthoryear{Karypis and Kumar}{Karypis and Kumar}{1997}]%
        {Karypis1998a}
\bibfield{author}{\bibinfo{person}{G. Karypis} {and} \bibinfo{person}{V.
  Kumar}.} \bibinfo{year}{1997}\natexlab{}.
\newblock \showarticletitle{{Multilevel k -way Partitioning Scheme for
  Irregular Graphs}}.
\newblock \bibinfo{journal}{{\it J. Parallel and Distrib. Comput.}}
  \bibinfo{volume}{47}, \bibinfo{number}{2} (\bibinfo{year}{1997}),
  \bibinfo{pages}{109--124}.
\newblock


\bibitem[\protect\citeauthoryear{Kernighan and Lin}{Kernighan and Lin}{1970}]%
        {Kernighan1970}
\bibfield{author}{\bibinfo{person}{B. Kernighan} {and} \bibinfo{person}{S.
  Lin}.} \bibinfo{year}{1970}\natexlab{}.
\newblock \showarticletitle{{An efficient heuristic procedure for partitioning
  graphs}}.
\newblock \bibinfo{journal}{{\em Bell systems technical journal\/}}
  \bibinfo{volume}{49}, \bibinfo{number}{2} (\bibinfo{year}{1970}),
  \bibinfo{pages}{291--307}.
\newblock


\bibitem[\protect\citeauthoryear{Lee, Moon, Park, et~al\mbox{.}}{Lee
  et~al\mbox{.}}{2008}]%
        {Lee2008}
\bibfield{author}{\bibinfo{person}{S. Lee}, \bibinfo{person}{B. Moon},
  \bibinfo{person}{C. Park}, {et~al\mbox{.}}} \bibinfo{year}{2008}\natexlab{}.
\newblock \showarticletitle{{A case for flash memory ssd in enterprise database
  applications}}. In \bibinfo{booktitle}{{\em Proc. SIGMOD}}.
  \bibinfo{pages}{1075}.
\newblock
\showISBNx{9781605581026}
\showISSN{01635999}


\bibitem[\protect\citeauthoryear{Lidl and Niederreiter}{Lidl and
  Niederreiter}{1997}]%
        {Lidl1997a}
\bibfield{author}{\bibinfo{person}{R. Lidl} {and} \bibinfo{person}{H.
  Niederreiter}.} \bibinfo{year}{1997}\natexlab{}.
\newblock \showarticletitle{{Finite Fields}}.
\newblock \bibinfo{journal}{{\em Encyclopedia of Mathematics and Its
  Applications\/}} (\bibinfo{year}{1997}), \bibinfo{pages}{1983}.
\newblock
\showISBNx{0-201-13519-1}


\bibitem[\protect\citeauthoryear{Margo and Seltzer}{Margo and Seltzer}{2015}]%
        {Margo15}
\bibfield{author}{\bibinfo{person}{D. Margo} {and} \bibinfo{person}{M.
  Seltzer}.} \bibinfo{year}{2015}\natexlab{}.
\newblock \showarticletitle{{A scalable distributed graph partitioner}}.
\newblock \bibinfo{journal}{{\em Proc. VLDB\/}} \bibinfo{volume}{8},
  \bibinfo{number}{12} (\bibinfo{year}{2015}), \bibinfo{pages}{1478--1489}.
\newblock
\showISBNx{9781450323789}
\showISSN{21508097}


\bibitem[\protect\citeauthoryear{McKay}{McKay}{1981}]%
        {McKay1981a}
\bibfield{author}{\bibinfo{person}{B. McKay}.} \bibinfo{year}{1981}\natexlab{}.
\newblock \bibinfo{title}{{Practical graph isomorphism}}.
\newblock   (\bibinfo{year}{1981}), \bibinfo{numpages}{45--87}~pages.
\newblock
\showISBNx{0747-7171}
\showISSN{07477171}


\bibitem[\protect\citeauthoryear{Mendelzon and Wood}{Mendelzon and
  Wood}{1995}]%
        {mendelzon1995finding}
\bibfield{author}{\bibinfo{person}{A. Mendelzon} {and} \bibinfo{person}{P.
  Wood}.} \bibinfo{year}{1995}\natexlab{}.
\newblock \showarticletitle{{Finding Regular Simple Paths in Graph Databases}}.
\newblock \bibinfo{journal}{{\it SIAM J. Comput.}} \bibinfo{volume}{24},
  \bibinfo{number}{6} (\bibinfo{year}{1995}), \bibinfo{pages}{1235--1258}.
\newblock
\showISBNx{1558601015}
\showISSN{0097-5397}


\bibitem[\protect\citeauthoryear{Moreau, Missier, Belhajjame,
  et~al\mbox{.}}{Moreau et~al\mbox{.}}{2012}]%
        {w3c-prov-dm}
\bibfield{author}{\bibinfo{person}{L. Moreau}, \bibinfo{person}{P. Missier},
  \bibinfo{person}{K. Belhajjame}, {et~al\mbox{.}}}
  \bibinfo{year}{2012}\natexlab{}.
\newblock \bibinfo{booktitle}{{\em {PROV-DM: The PROV Data Model}}}.
\newblock \bibinfo{type}{{T}echnical {R}eport}. \bibinfo{institution}{World
  Wide Web Consortium}.
\newblock


\bibitem[\protect\citeauthoryear{Nishimura and Ugander}{Nishimura and
  Ugander}{2013}]%
        {Nishimura2013}
\bibfield{author}{\bibinfo{person}{J. Nishimura} {and} \bibinfo{person}{J.
  Ugander}.} \bibinfo{year}{2013}\natexlab{}.
\newblock \showarticletitle{{Restreaming graph partitioning}}. In
  \bibinfo{booktitle}{{\em Proc. SIGKDD}}. \bibinfo{address}{New York, New
  York, USA}, \bibinfo{pages}{1106--1114}.
\newblock
\showISBNx{9781450321747}


\bibitem[\protect\citeauthoryear{Pavlo, Curino, and Zdonik}{Pavlo
  et~al\mbox{.}}{2012}]%
        {Pavlo2012}
\bibfield{author}{\bibinfo{person}{A. Pavlo}, \bibinfo{person}{C. Curino},
  {and} \bibinfo{person}{S. Zdonik}.} \bibinfo{year}{2012}\natexlab{}.
\newblock \showarticletitle{{Skew-aware automatic database partitioning in
  shared-nothing, parallel OLTP systems}}. In \bibinfo{booktitle}{{\em Proc.
  SIGMOD}}. \bibinfo{pages}{61}.
\newblock
\showISBNx{9781450312479}
\showISSN{07308078}


\bibitem[\protect\citeauthoryear{Peng, Zou, Chen, et~al\mbox{.}}{Peng
  et~al\mbox{.}}{2016}]%
        {Peng2016}
\bibfield{author}{\bibinfo{person}{P. Peng}, \bibinfo{person}{L. Zou},
  \bibinfo{person}{L. Chen}, {et~al\mbox{.}}} \bibinfo{year}{2016}\natexlab{}.
\newblock \showarticletitle{{Query Workload-based RDF Graph Fragmentation and
  Allocation}}. In \bibinfo{booktitle}{{\em Proc. EDBT}}.
  \bibinfo{pages}{377--388}.
\newblock


\bibitem[\protect\citeauthoryear{Pujol, Erramilli, Siganos, Yang, Laoutaris,
  Chhabra, and Rodriguez}{Pujol et~al\mbox{.}}{2010}]%
        {Pujol2010a}
\bibfield{author}{\bibinfo{person}{Josep~M Pujol}, \bibinfo{person}{Vijay
  Erramilli}, \bibinfo{person}{Georgos Siganos}, \bibinfo{person}{Xiaoyuan
  Yang}, \bibinfo{person}{Nikos Laoutaris}, \bibinfo{person}{Parminder
  Chhabra}, {and} \bibinfo{person}{Pablo Rodriguez}.}
  \bibinfo{year}{2010}\natexlab{}.
\newblock \showarticletitle{{The little engine(s) that could}}. In
  \bibinfo{booktitle}{{\em Proc. SIGCOMM}}. \bibinfo{pages}{375--386}.
\newblock
\showISBNx{9781450302012}


\bibitem[\protect\citeauthoryear{Quamar, Kumar, and Deshpande}{Quamar
  et~al\mbox{.}}{2013}]%
        {Quamar2013}
\bibfield{author}{\bibinfo{person}{A. Quamar}, \bibinfo{person}{K. Kumar},
  {and} \bibinfo{person}{A. Deshpande}.} \bibinfo{year}{2013}\natexlab{}.
\newblock \showarticletitle{{SWORD}}. In \bibinfo{booktitle}{{\em Proc. EDBT}}.
  \bibinfo{pages}{430}.
\newblock
\showISBNx{9781450315975}


\bibitem[\protect\citeauthoryear{Ribeiro and Silva}{Ribeiro and Silva}{2014}]%
        {Ribeiro2010}
\bibfield{author}{\bibinfo{person}{P. Ribeiro} {and} \bibinfo{person}{F.
  Silva}.} \bibinfo{year}{2014}\natexlab{}.
\newblock \showarticletitle{{G-Tries: a data structure for storing and finding
  subgraphs}}.
\newblock \bibinfo{journal}{{\em Data Mining and Knowledge Discovery\/}}
  \bibinfo{volume}{28}, \bibinfo{number}{2} (\bibinfo{year}{2014}),
  \bibinfo{pages}{337--377}.
\newblock
\showISBNx{9781605586380}
\showISSN{1384-5810}


\bibitem[\protect\citeauthoryear{Shang and Yu}{Shang and Yu}{2013}]%
        {Shang2013}
\bibfield{author}{\bibinfo{person}{Z. Shang} {and} \bibinfo{person}{J. Yu}.}
  \bibinfo{year}{2013}\natexlab{}.
\newblock \showarticletitle{{Catch the Wind: Graph workload balancing on
  cloud}}.
\newblock \bibinfo{journal}{{\em Proc. International Conference on Data
  Engineering (ICDE)\/}} (\bibinfo{year}{2013}), \bibinfo{pages}{553--564}.
\newblock
\showISBNx{978-1-4673-4910-9}


\bibitem[\protect\citeauthoryear{Song, Ge, Chen, and Wang}{Song
  et~al\mbox{.}}{2014}]%
        {Song2014}
\bibfield{author}{\bibinfo{person}{C. Song}, \bibinfo{person}{T. Ge},
  \bibinfo{person}{C. Chen}, {and} \bibinfo{person}{J. Wang}.}
  \bibinfo{year}{2014}\natexlab{}.
\newblock \showarticletitle{{Event pattern matching over graph streams}}.
\newblock \bibinfo{journal}{{\em Proc. VLDB\/}} \bibinfo{volume}{8},
  \bibinfo{number}{4} (\bibinfo{year}{2014}), \bibinfo{pages}{413--424}.
\newblock
\showISSN{21508097}


\bibitem[\protect\citeauthoryear{Stanton and Kliot}{Stanton and Kliot}{2012}]%
        {Stanton2012}
\bibfield{author}{\bibinfo{person}{I. Stanton} {and} \bibinfo{person}{G.
  Kliot}.} \bibinfo{year}{2012}\natexlab{}.
\newblock \showarticletitle{{Streaming graph partitioning for large distributed
  graphs}}. In \bibinfo{booktitle}{{\em Proc. SIGKDD}}.
  \bibinfo{pages}{1222--1230}.
\newblock
\showISBNx{978-1-4503-1462-6}


\bibitem[\protect\citeauthoryear{Tsourakakis, Gkantsidis, Radunovic,
  et~al\mbox{.}}{Tsourakakis et~al\mbox{.}}{2014}]%
        {Tsourakakis2012}
\bibfield{author}{\bibinfo{person}{C. Tsourakakis}, \bibinfo{person}{C.
  Gkantsidis}, \bibinfo{person}{B. Radunovic}, {et~al\mbox{.}}}
  \bibinfo{year}{2014}\natexlab{}.
\newblock \showarticletitle{{FENNEL}}. In \bibinfo{booktitle}{{\em Proc. ACM
  International conference on Web search and data mining (WSDM)}}.
  \bibinfo{pages}{333--342}.
\newblock
\showISBNx{9781450323512}


\bibitem[\protect\citeauthoryear{Xu, Chen, and Cui}{Xu et~al\mbox{.}}{2014}]%
        {Xu2014}
\bibfield{author}{\bibinfo{person}{N. Xu}, \bibinfo{person}{L. Chen}, {and}
  \bibinfo{person}{B. Cui}.} \bibinfo{year}{2014}\natexlab{}.
\newblock \showarticletitle{{LogGP}}.
\newblock \bibinfo{journal}{{\em Proc. VLDB\/}} \bibinfo{volume}{7},
  \bibinfo{number}{14} (\bibinfo{year}{2014}), \bibinfo{pages}{1917--1928}.
\newblock
\showISSN{21508097}


\end{thebibliography}

\end{document}